\documentclass[floatfix,prc,twocolumn,showpacs,amsmath,amssymb]{revtex4}
\usepackage{epsfig}
\usepackage{color}
\usepackage{amssymb}
\usepackage{amsmath}
\usepackage[ngerman,english]{babel}
\usepackage{multirow}
\usepackage{graphicx}
\usepackage{subfigure}
\usepackage{natbib}

\begin{document}

\title{Employing ternary fission of $^{242}$Pu as a probe of very neutron rich  matter}

\author{J.B. Natowitz}
\affiliation{Cyclotron Institute, Texas A{\&}M University, College Station, Texas 778431}
\author{H. Pais}
\affiliation{CFisUC, Department of Physics, University of Coimbra, 3004-516 Coimbra, Portugal}
\author{G. R{\"o}pke}
\affiliation{Institute of Physics, University of Rostock, 18059 Rostock, Germany}

\date{\today}

\begin{abstract}
Detailed assessments of the ability of recent theoretical approaches to modeling existing experimental data for ternary fission confirm earlier indications that the dominant mode of cluster formation in ternary fission is clusterization in very neutron rich, very low density, essentially chemically equilibrated, nucleonic matter. An extended study and comparison of these approaches applied to ternary fission yields in the thermal neutron induced reaction $^{241}$Pu($n_{\rm th}$,f) has been undertaken to refine the characterization of the source matter. The resonance gas approximation has been improved taking in-medium effects on the binding energies into account. A temperature of 1.29 MeV, density of $6.7 \times 10^{-5}$ nucleons/fm$^3$ and proton fraction $Y_p$ = 0.035 are found to provide a good representation of yields of the ternary emitted  light particles and clusters. In particular, results for $Z= 1$ and 2 isotopes are presented. Isotopes with larger $Z$ are discussed, and the roles of medium and continuum effects, even at very low density are illustrated. \end{abstract}


\maketitle

\section{Introduction}
In the neutron induced or spontaneous ternary fission of a heavy isotope a nucleon or light cluster is emitted perpendicular to the fission axis determined by the two separating large fission fragments \cite{Halp,Wage, 
Mills,IAEA96,Krappe,Shub, Vorob,Gusev,Koes1,Koes2,Kopatch}. 
$^4$He, emitted in approximately 1/500 events is the most dominant charged particle
but other charged isotopes with charge number $Z=1$ up to $Z=18$ have been observed \cite{ Koes1,Koes2}. These light charged particles (LCP) are emitted from the neck region at the time of scission and may be considered as signals which describe the state of nuclear matter in the neck at that time.
 
To interpret the observed ternary yields, statistical models have been applied which assume thermodynamic equilibrium at chemical freeze-out during scission \cite{ Valskii2, Lestone08, Sara}. However, experimental yields for the heavier elements are typically overestimated unless some mechanism for suppression of higher mass products is introduced \cite{Lestone08, Sara}. In Wuenschel {\it et al.} \cite{Sara}, chemical equilibrium is achieved in accordance with the grand canonical ensemble only for the lightest isotopes and a time dependent nucleation process for production of heavier LCP is introduced \cite{Schmelzer,Demo} so that the LCP yield becomes increasingly suppressed with increasing mass number.

Recently, this approach has been explored in more detail and used to determine isotopic equilibrium constants for LCP emitted in the ternary fission reaction $^{241}$Pu($n_{\rm th}$,f) \cite{natowitz20}. Further investigations were then undertaken to better characterize the ternary fissioning $^{242}$Pu source. In \cite{rnp20}, the simple ideal model of nuclear statistical equilibrium was improved considering medium effects and continuum correlations \cite{R20}, e.g., resonances such as $^4$H, $^5$He, $^8$Be (as known from the virial expansion of the nuclear matter equation of state).
A nearly perfect description of the measured yields of H and He isotopes for $^{252}$Cf(sf) was obtained. 
In \cite{rnp21}, several different fission reactions were investigated within an information entropy approach, and isotopes up to $Z=6$ are included.
These investigations showed that the dominant mode of cluster formation in ternary fission is clusterization in very neutron rich, very low density, essentially chemically equilibrated, nucleonic matter at temperatures of about 1 MeV.

In the present work, an extended study and comparison of the different approaches applied to modeling ternary fission yields in the thermal neutron induced reaction $^{241}$Pu($n_{\rm th}$,f) \cite{Koes1, Koes2} is undertaken. A more detailed exploration of the in-medium and continuum effects leads to a more refined characterization of the source matter.  
We show that, in particular, the weakly bound states are strongly influenced by in-medium effects and provide an alternative observable to determine the density of the source matter. We find that the neck matter at scission has a temperature $T\approx 1.29$ MeV, a nucleon density $n_B \approx 6.7 \times 10^{-5}$ nucleons/fm$^3$, and a proton fraction $Y_p \approx 0.035$. 
This work provides a baseline laboratory test of the low-density nuclear equation of state for conditions which may be encountered in astrophysical sites such as core-collapse supernova events in which a neo-neutron star is formed, evolves with time and cools down to a neutron star \cite{Bezn20}. Similar neutron rich matter may also be produced in the merging of a binary neutron star system. Both scenarios   relax to a temperature of about 1 MeV in less than a minute. In such dense, highly excited systems, surface densities of $10^{11}$ g/cm$^3$ are typical, but the proton fraction may be different. In $\beta$ equilibrium, it depends upon the neutrino density, see, e.g., \cite{Arcones08,Fischer}. 
Even if the proton fraction in astrophysical scenarios is different from the proton fraction in the neck matter at scission, the equations of state (for a review see, e.g., \cite{Oertel17,Furusawa22}) employed over a wide region of parameter values may be checked at the special conditions of scission. Investigations of the correct description of correlations under scission conditions may also be applied to other regions of the astrophysical parameter space.

\section{Yields from a quantum statistical approach}

We consider different stages of the fission process. For the time evolution of the nucleonic system up to scission we assume a quick relaxation to local thermodynamic equilibrium. Within the statistical model framework the grand canonical distribution at scission, where chemical freeze-out takes place, gives the primordial (or primary) yields for the relevant species. For the time evolution after scission we assume a reaction kinetics in which populated excited states decay, 
and the kinetic energies are determined by the interaction between the fission products. The decay of unstable nuclear states and resonances is described as feed-down processes which transform the primordial distribution of yields to the final observable yield distribution. 
This approach to the time evolution of the fission process may be considered as an approximation within the systematic approach of 
non-equilibrium statistical operators where both stages of time evolution, the hydrodynamical and kinetic ones, are unified within an information theoretical approach \cite{rnp20,rnp21}. This information theoretical approach allows us to introduce Lagrange parameters $\lambda_T(t), \lambda_n(t), \lambda_p(t)$ 
which are the nonequilibrium generalizations of the temperature and the chemical potentials of neutrons ($n$) and protons ($p$).
They depend on time $t$ and, in general for the hydrodynamical description, also on position.

In the present work, we focus on inferring the corresponding Lagrange parameters $\lambda_i$ for the primary distribution at chemical freeze-out from the observed final yield distribution. For this, we need an accurate solution of the grand canonical distribution at scission. We employ and compare three successive approximations:\\
(i) The resonance gas approximation (res.gas) known also as nuclear statistical equilibrium (NSE), see \cite{Furusawa22,Ropke82} where the nucleonic system is considered as an ideal mixture of nuclei in the ground state and in (unstable) excited states and resonances. A semi-empirical improvement is the excluded volume model \cite{Hemp1,Hemp2}.\\
(ii) The virial approximation (vir) where binary interactions between the different constituents are taken into account considering the respective scattering phase shifts  \cite{Ropke82,SRS,HS}.\\
(iii) Accounting for in-medium corrections (medium) such as self-energy shifts and Pauli blocking effects \cite{Ropke82,R09,Typel10}. 

In particular, we investigate whether the frequently used nuclear statistical equilibrium model is sufficient to describe nucleonic systems under the scission conditions in the neck region or whether continuum correlations and in medium effects must be taken into account. 

In the quantum statistical approach, after the cluster decomposition of the spectral function, the density is decomposed into partial densities of different channels characterized by $A,Z$ \cite{rnp20}. 
The primordial yields, here denoted as relevant yields $Y^{\rm rel, approx}_{A,Z}$ in the corresponding approximation are calculated as 
\begin{eqnarray}
\label{Y0}
&&Y^{\rm rel, approx}_{A,Z}  \propto  R^{\rm approx}_{A,Z}(\lambda_T)\, g_{A,Z} \left(\frac{2 \pi \hbar^2}{A m \lambda_T}\right)^{-3/2}  \nonumber \\
&& \times e^{(B_{A,Z}+(A-Z) \lambda_n+Z  \lambda_p)/ \lambda_T},
\end{eqnarray}  
where $B_{A,Z}$ denotes the (ground state) binding energy and $g_{A,Z}$ the degeneracy \cite{Nudat}.
The prefactor
\begin{equation} 
\label{Rappr}
R^{\rm approx}_{A,Z}(\lambda_T)=1+\sum^{\rm exc}_i [g_{AZ,i}/g_{A,Z}] e^{-E_{AZ,i}/\lambda_T}
\end{equation}  
is related to the intrinsic partition function of the cluster $\{A,Z\}$. 
The summation is performed over all excited states $i$, excitation energy $E_{AZ,i}$ and degeneracy $g_{AZ,i}$  \cite{Nudat}.

Different approximations are considered for the intrinsic partition function as discussed above.\\
(i) In the the resonance gas approximation, the summation for $R^{\rm res.gas}_{A,Z}(\lambda_T)$ is performed over all states including resonances as found in the nuclear data tables \cite{Nudat}. The angular momentum degeneracies are included.\\
(ii) 
In the virial approximation, $R^{\rm vir}_{A,Z}(\lambda_T)$ includes also the sum over all scattering states (continuum correlations) as expressed by the Beth-Uhlenbeck formula. Expressions and interpolation formulas are found in \cite{rnp20,R15}. \\
(iii) 
The expression  $R^{\rm,medium}_{A,Z}(\lambda_T,\lambda_n,\lambda_p)$ takes in-medium effects into account, in particular the shifts of binding energy values because of self-energy, Pauli blocking effects and the modification of bound state wave functions and scattering phase shifts. Therefore, this term is also dependent on the densities of neutrons and protons.
Expressions for the intrinsic partition function and $R^{\rm res.gas}_{A,Z}, R^{\rm vir}_{A,Z}, R^{\rm medium}_{A,Z}$ for $\lambda_T=1.29 \,{\rm MeV}$ are given for $Z \le 6$ in the appendix, Tabs. \ref{tab:He} - \ref{tab:C}, see also Sec. \ref{sec:3} below. 

The summation in (\ref{Rappr}) is performed over all excited states $i$ which are bound. Also, the continuum contributions are included in the virial expression. For instance, 
the Beth-Uhlenbeck formula expresses the contribution of the continuum to the intrinsic partition function 
via the scattering phase shifts, see \cite{SRS,HS,R20,R09,R11,R15}. 
The simple statistical equilibrium distribution (NSE) is obtained for $R^{\rm vir}_{A,Z}(\lambda_T)=1$, 
i.e., neglecting the contribution of all excited states including continuum correlations.

As an example we focus on the fission reaction $^{241}$Pu($n_{\rm th}$,f)  induced by thermal neutrons where good data for the ternary fission yields are available \cite{Koes1, Koes2}. 
The observed yields up to $^{20}$C are shown in Tabs. \ref{241Pujun1}, \ref{241Pujun2} below. Instead of normalizing to the yield of total $\alpha$ particle emission, assigned to be 10000, we employ absolute yields per fission. The total $\alpha$ particle emission yield per fission was taken from \cite{IAEA96}.

The Koester data do not contain values for scission nucleon ($n$ or $p$) emission. Determination of those yields is experimentally very challenging \cite{Vorob, Gusev, Shub}.
Experimental scission proton yields are very small and careful experiments have revealed that secondary processes dominate the apparent yields reported \cite{Shub}. In our opinion the best available constraint on the scission proton yield is that of reference \cite{Shub}, where only an upper limit of 2.9 to $4.0 \times 10^{-5}$ is deduced. Interestingly, although the determination of scission neutron emission yield in the presence of a much larger yield of secondary neutrons evaporated from the separated fission fragments is inherently even more difficult, very precise measurements and analyses of the neutron energy and angular distributions have been carried out and lead to the conclusion that the scission neutron yield is 0.107 per fission, approximately one thirtieth of the total neutron yield \cite{Vorob, Gusev}. We have adopted this number for our analysis. It is worth noting that, in an equilibrium picture, the relative scission neutron and proton yields implied by these results indicate that the neck matter at scission is extremely neutron rich. That this must be the case has previously been inferred from the fact that $^3$He has not been detected in ternary fission experiments while the isotope $^3$H, with a similar binding energy is the second most abundant ternary charged fragment observed \cite{Mills, IAEA96}.

 \begin{table*}
\begin{center}
\hspace{0.5cm}
 \begin{tabular}{|c|c|c|c|c|c|c|c|}
\hline
isotope& $A$  &  $Z$ & $Y^{\rm obs}_{A,Z}$&   $R_{A,Z}^{\rm res.gas}(1.29)$   & $Y_{A,Z}^{\rm rel,res.gas}$ &$Y_{A,Z}^{\rm final,res.gas}$ &$X^{\rm res.gas}_{A,Z}$\\
\hline
1n		&1&0		& 0.107 		&1			&0.1098  		&0.1098 			&0.9215 \\
1H		&1& 1 		& -      			& 1			&4.298E-6   &4.298E-6 		&0.00003414  \\
2H 		&2& 1 		&8.463E-6 	&	1			&8.889E-6 	&8.889E-6  		&0.000146  \\
3H		&3& 1 		&1.584E-4 	&	1			&1.212E-4	&1.398E-4 		&0.00359   \\
4H*		&4& 1 		& -     			&	1.579	&1.855E-5	& [$\to ^3$H]  &-  \\
\hline	
3He		&3& 2 		& -     			& 1			& 2.549E-9	& 2.817E-9   	&6.163E-8  \\
4He		&4& 2 		&2.015E-3	& 1      		&1.449E-3  	&1.911E-3		& 0.06529\\
5He*	&5& 2 		& -     			& 1			&3.999E-4   &[$\to ^4$He] &- \\
6He$^0$&6&2	&5.239E-5	& 1	    	&4.322E-5   &5.916E-5		&0.003232 \\
6He*	&6& 2 		& -     			& 1.242	&5.407E-5   &[$\to ^4$He] &- \\
7He*	&7& 2 		& -     			& 1.156  	& 1.594E-5  &[$\to ^6$He]& -  \\
8He		&8& 2 		&3.022E-6 	& 1		  	&2.605E-6  	&2.898E-6		& 0.0002249 \\
8He*	&8& 2 		& -     			& 0.452  	&1.193E-6  	&[$\to ^4$He] & -\\
9He*	&9& 2 		& -     			& 1.426  	&2.929E-7   &[$\to ^8$He]&- \\
\hline
6Li		&6 & 3 	& -    			& 1			& 4.059E-8  &4.059E-8 		&2.085E-6  \\
6Li*		&6 & 3 	& -    			& 0.5471	& 2.46E-8    &[$\to ^3$H]	&-  \\
7Li		&7 & 3 	&1.35E-6		& 1.345   &2.207E-6 	&2.545E-6		&0.0001591  \\
8Li		&8 & 3 	&8.463E-7 	& 1.281   &1.357E-6   &1.357E-6  		&0.00009983  \\
8Li*		&8 & 3 	& -  				&0.3151  	& 3.374E-7  &[$\to ^7$Li]	& -  \\
9Li		&9 & 3 	&1.672E-6 	& 1.062   	&2.181E-6 	&2.335E-6 		&0.0002005  \\
10Li*	&10& 3 	& -    			& 1			&1.531E-7   &[$\to ^9$Li]	&-\\
11Li$^0$	&11& 3&9.068E-10	& 1			&2.785E-8 	&2.962E-8 		&3.312E-6  \\
12Li*	&12& 3	& -    			& 1			& 1.773E-9  &[$\to ^{11}$Li] &- \\
\hline
7Be		&7  & 4 	& -    			& 1.359    &2.382E-11 &2.382E-11    &1.401E-9 \\
8Be*		&8  & 4 	& -    			& 1.477	&1.59E-6      & [$\to ^4$He] &-\\
9Be		&9  & 4 	&8.866E-7 	& 1   		&1.616E-6 	&1.616E-6 		&0.0001325  \\
9Be*		&9  & 4 	& - 				& 0.5628  &1.244E-6 	&[$\to ^4$He]  &-  \\
10Be	&10 & 4 	&9.269E-6	& 1.367    &1.112E-5 	&1.51E-5 		&0.001435  \\
10Be$^0$&10&4& -    			& 0.0789	& 6.553E-7  &[$\to ^{10}$Be]  &- \\
11Be$^0$&11&4&1.189E-6 	& 1    		&2.422E-6	&4.313E-6 		&0.0004634  \\
11Be$^0$&11&4& -		 		& 0.7803 &1.891E-6	&[$\to ^{11}$Be] &-   \\
11Be*	&11 & 4	& -    			& 1.343	& 3.287E-6  & [$\to ^{10}$Be]& -  \\
12Be	&12 & 4	&5.642E-7	& 2.149  	&2.781E-6   &3.73E-6  		&0.0004532 \\
12Be$^0$&12&4& -    			& 0.3657	&7.655E-7 	& [$\to ^{12}$Be] & -  \\
13Be*	&13 & 4	& -    			& 1			& 1.84E-7	& [$\to ^{12}$Be]    & - \\
14Be	&14 & 4	&5.441E-10 & 1  			&3.606E-8  	&4.117E-8 		&6.237E-6   \\
15Be*	&15 & 4	& -    			& 1			& 5.114E-9	& [$\to ^{14}$Be]& - \\
\hline
 \end{tabular}
\caption{Observed yields per fission of ternary fission of $^{241}$Pu($n_{\rm th}$,f), including 0.107 for scission neutrons (col. 4), are compared to a final state distribution, calculated in the resonance gas approximation. The Lestone fit metric \cite{Lestone08}  is calculated for the isotopes with $Z \le 2$. The minimum is found for the parameter values given at the end of table. $Y_{A,Z}^{\rm rel,res.gas}$ denotes primordial yield, $Y_{A,Z}^{\rm final,res.gas}$: final yield, $X^{\rm res.gas}_{A,Z}$: mass fraction. Units: MeV, fm.}
\label{241Pujun1}
\end{center}
\end{table*}

 \begin{table*}
\begin{center}
\hspace{0.5cm}
 \begin{tabular}{|c|c|c|c|c|c|c|c|c|c|c|c|}
\hline
isotope& $A$  &  $Z$ & $Y^{\rm obs}_{A,Z}$&   $R_{A,Z}^{\rm res.gas}(1.29)$   & [$Y_{A,Z}^{\rm rel,res.gas}$ ]&[$Y_{A,Z}^{\rm final,res.gas}$] &[$X^{\rm res.gas}_{A,Z}$]\\
\hline
10B	&10 & 5 			&  -  				&  1.363     	&2.484E-9 		& 2.565E-9    			&2.295E-7  \\
10B$^0$	&10 & 5 	&  -  				&  0.0443    &8.17E-11 		& [$\to ^{10}$B]    &-   \\
11B	&11 & 5 			&3.224E-7	& 1.175    	& 8.876E-7   	&8.876E-7 				&0.00009106 \\
12B	&12 & 5 			&2.015E-7	& 2.251   		&1.703E-6   	&1.835E-6  				&0.0002126  \\
12B$^0$&12 & 5 	&  -  				&  0.1715 	& 1.32E-7        &  [$\to ^{12}$B] & -  \\
13B	&13 & 5 			&  -  				& 1    			& 4.367E-6      & 4.91E-6 				&0.0006411 \\
14B$^0$	&14 & 5 	&2.62E-8 	&1     			&1.123E-6  		&1.504E-6  				& 0.0002177  \\
14B$^0$&14 & 5 	&  -  				& 0.3381 		& 3.809E-7   	&  [$\to ^{14}$B]  & - \\
14B*&14 & 5 			&  -  				& 0.4803 		& 5.426E-7  	 &  [$\to ^{13}$B]  & -  \\
15B	&15 & 5 			&9.269E-9 	& 1   			&7.523E-7   	&7.691E-7 				&0.0001235  \\
16B*	&16 & 5 		&  -  				& 1  				&1.685E-8   	& [$\to ^{15}$B]   & -  \\
17B	&17 & 5 			&  -  				& 1   	 		&2.021E-8		& 2.26E-8 				& 4.397E-6 \\
18B*	&18 & 5 		&  -  				& 1 				& 2.389E-9      &  [$\to ^{17}$B]& -  \\
\hline
13C	&13 & 6 			&  -  				& 1.358    	& 2.079E-6      &2.079E-6   			& 0.0002587 \\
14C	&14 & 6 			&2.539E-6 	& 1.069   		&4.479E-5 		&5.129E-5 				&0.007177  \\
15C	&15 & 6  			&8.665E-7 	& 1    			&2.08E-5  		&5.606E-5				&0.008647  \\
15C$^0$&15 & 6 	&  -  				& 1.69			&3.527E-5       & [$\to ^{15}$C] & -  \\
15C*&15 & 6 			&  -  				& 0.3074		&6.501E-6       & [$\to ^{14}$C] & -  \\
16C	&16 & 6 			&1.008E-6 	& 2.272    	&6.149E-5 		&9.799E-5 				&0.01218  \\
16C$^0$&16 & 6 	&  -  				& 0.885 		& 2.425E-5	   	& [$\to ^{16}$C]    & -    \\
17C$^0$	&17 & 6 	&1.29E-7 	& 1    			&1.816E-5  		&4.693E-5  				&0.008776  \\
17C$^0$	&17 & 6 	&-	 				& 1.582   		&2.877E-5  		&[$\to ^{17}$C]  	&-  \\
17C*&17 & 6 			&  -  				& 0.6677		&1.225E-5	   	&[$\to ^{16}$C]   	& -    \\
18C	&18 & 6 			&5.642E-8  	&  3.172    	&3.515E-5   	&4.42E-5 				&0.009127  \\
19C$^0$	&19 & 6 	&5.038E-10	& 5.112    	& 1.662E-5 		&1.662E-5				 &0.003715   \\
19C*&19 & 6 			&  -  				& 2.776 		& 9.051E-6	   	& [$\to ^{18}$C]   & -      \\
20C	&20 & 6 			&7.254E-10	& 2.426   		& 3.743E-6  	& 3.743E-6 				&0.0009117  \\
\hline
\hline
$\lambda_T$&   & & &1.2897  & & &  \\
$\lambda_n$&   & & &-3.1486 & & &  \\
$\lambda_p$&   & & &-16.273 & & &  \\
\hline
volume  &   & & &1859.4& & &  \\
$n_B$   &   & & &0.000064 & & &   \\
$Y_p$  &   & & &0.03486 & & &  \\
 \hline
fit metric &   & & &0.005485 & & &   \\
\hline
 \end{tabular}
\caption{Continuation of Tab. \ref{241Pujun1}. Observed yields per fission of ternary fission of $^{241}$Pu($n_{\rm th}$,f) (col. 4) are compared to a final state distribution, calculated in resonance gas approximation, including 0.107 for scission neutrons. The Lestone fit metric is calculated for the isotopes with $Z \le 2$. The minimum is found for the parameter values given at the end of table.  Units: MeV, fm.}
\label{241Pujun2}
\end{center}
\end{table*}

We are interested in a quantum statistical description of dense matter. The standard nuclear statistical equilibrium (NSE) approach describes matter in the low-density limit where interaction between the components may be neglected (ideal gas of nucleons and nuclei in ground and excited states). We are interested in a consistent quantum statistical description of interacting components as done, for instance, for the equation of state \cite{Qin12}. In particular we take into account continuum correlations and in-medium effects.\\

\section{Extraction of Lagrange parameters from observed yields}
\label{sec:3}

In this section we demonstrate how the Lagrange parameters $\lambda_i$
may be extracted from the observed yields. For simplicity, we use in this example only the simplest approximation for the intrinsic partition function, i.e., the
ideal resonance gas approximation. $R_{A,Z}^{\rm
res.gas}(\lambda_T)$ is calculated according (\ref{Rappr}) where for isotopes
$\{A,Z\}$ all excited states $i$ are summed over, the excitation
energies $E_{AZ,i}$ and angular momentum degeneracies
$g_{AZ,i}=2J_{AZ,i}+1$ are taken from the nuclear data tables
\cite{Nudat}. In the Appendix we give these values for $Z \le 6$ as well as the ground state 
  binding energies and the threshold
energies for the continuum of scattering states $E_{A,Z}^{\rm thresh}$. In
Tabs. \ref{tab:He} - \ref{tab:C}, we divide the intrinsic partition function into
different parts with respect to the contribution to the final yields:
The summation over all excited states above the threshold energy is
marked by an asterisk. After freeze-out, these excited states are
assumed to decay and feed the yields of daughter
isotopes observed in the final distribution. This process is also explained in
 Tabs. \ref{241Pujun1}, \ref{241Pujun2}, where in col. 7 the respective feed-down channel
is indicated.
The other excited states are assumed to gamma decay to the ground state and
remain in the same isotope channel.
In Tabs. \ref{tab:He} - \ref{tab:C}, we indicate by the superscript "0" the states
which have a binding energy smaller than 1 MeV below the continuum
threshold.
These weakly bound states are of special interest when in-medium effects are considered below in Sec. \ref{inmedium}.

We use this subdivision of the intrinsic partition function to model the
kinetic stage of evolution, i.e. the transition from the primary
distribution to the final distribution considering only decay processes
of the excited nuclei. In a more general approach, this sharp subdivision
should be replaced by branching ratios describing the feed-down to the
final yields or using reaction networks. 

For the resonance gas approximation and
the virial approximation the $R$ factors are functions of the
temperature-like parameter $\lambda_T$. If in-medium effects are taken
into account, the $R$ factor depends in addition on the chemical potentials $\lambda
_n, \lambda _p$. Within our fit procedure described below, they are
 determined self-consistently. Values for the $R$ factors for
$\lambda_T=1.29$ MeV are given in the appendix, Tabs. \ref{tab:He} - \ref{tab:C},
for the resonance gas and virial approximations. In Tabs. \ref{241PuG}, \ref{241PuH}
values $R_{A,Z}^{\rm medium}(\lambda_T,\lambda_n,\lambda_p)$ are
shown for
$\lambda_T=1.29$ MeV, $\lambda_n=-3.09$ MeV, and $ \lambda_p=-16.19$ MeV, 
which take into account in-medium effects.
With a given set of Lagrange parameters, we are able to calculate the primary
yields and the final yields. This is shown for the resonance gas
approximation in Tabs. \ref{241Pujun1}, \ref{241Pujun2} as well as in
Tabs. \ref{241PuG}, \ref{241PuH} where in-medium effects are taken into account. 
In each case the calculated final yields are compared with the observed yields. 

The task is then to find values for the Lagrange parameters which
reproduce the observed yield in an optimum way.
We have previously observed \cite{Sara,rnp20,rnp21} that yields of isotopes with large mass number are suppressed because of nucleation kinetics or size effects.
For the fission process $^{241}$Pu($n_{\rm th}$,f) this suppression
with respect to the grand canonical distribution is observed for $A>10$.
To avoid this effect we use only the charged particles up to $Z=2$, i.e., the observed yields of $^1$n, $^2$H, $^3$H, $^4$He, $^6$He, $^8$He to find
values for the Lagrange parameters.

  As in Ref. \cite{Sara}, we use for the fit metric that of Lestone
\cite{Lestone08},
  defined by
  \begin{equation}
  \label{metric}
  M^2=\sum_{A,Z}^n \left(\ln[Y_{A,Z}^{\rm final,x}]-\ln[Y_{A,Z}^{\rm
obs}]\right)^2/n
  \end{equation}
  where $n$ is the number of fitted experimental data
  points.

  As shown in Tab.  \ref{241Pujun2}, the values $\lambda_T=1.29$ MeV, $\lambda_n=-3.149$ MeV, and $\lambda_n=-16.273$ MeV  are obtained from the
minimum of the fit metric for the resonance gas approximation.
The fit metric is also given in Tab.  \ref{241Pujun2}.  
From the yields per fission, we can also derive a volume.
The baryon density $n_B$ is obtained from the observed yields per fission $n_B=\sum_{A,Z} A Y^{\rm obs}_{A,Z}$ divided by the volume, 
it is mainly determined by the neutron density. 
The proton fraction $Y_p$ is obtained as $\sum_{A,Z} Z Y^{\rm obs}_{A,Z}/n_B$, it is mainly determined by the yield of $^4$He.
There is an uncertainty because the free proton density is not included in the fit, 
but the calculated values $Y_{1,1}$ are small so that no large effect is expected if the contribution of $^1$H is dropped.

The observed values $^1$n$^{\rm obs}$,  $^2$H$^{\rm obs}$, $^3$H$^{\rm obs}$, $^4$He$^{\rm
obs}$, $^6$He$^{\rm obs}$, and $^8$He$^{\rm obs}$ are well reproduced.
Below, in the next Section, we will also discuss the isotopes with $Z > 2$.

Whereas the values for temperature and the proton fraction are expected
and understood, the value of the optimum baryon density is to be determined. We
calculate the composition assuming different values for the density to
show that there is a strong variation of the composition and a sharp
minimum of the fit metric.

\begin{figure}[h]
\includegraphics[width=0.5\textwidth]{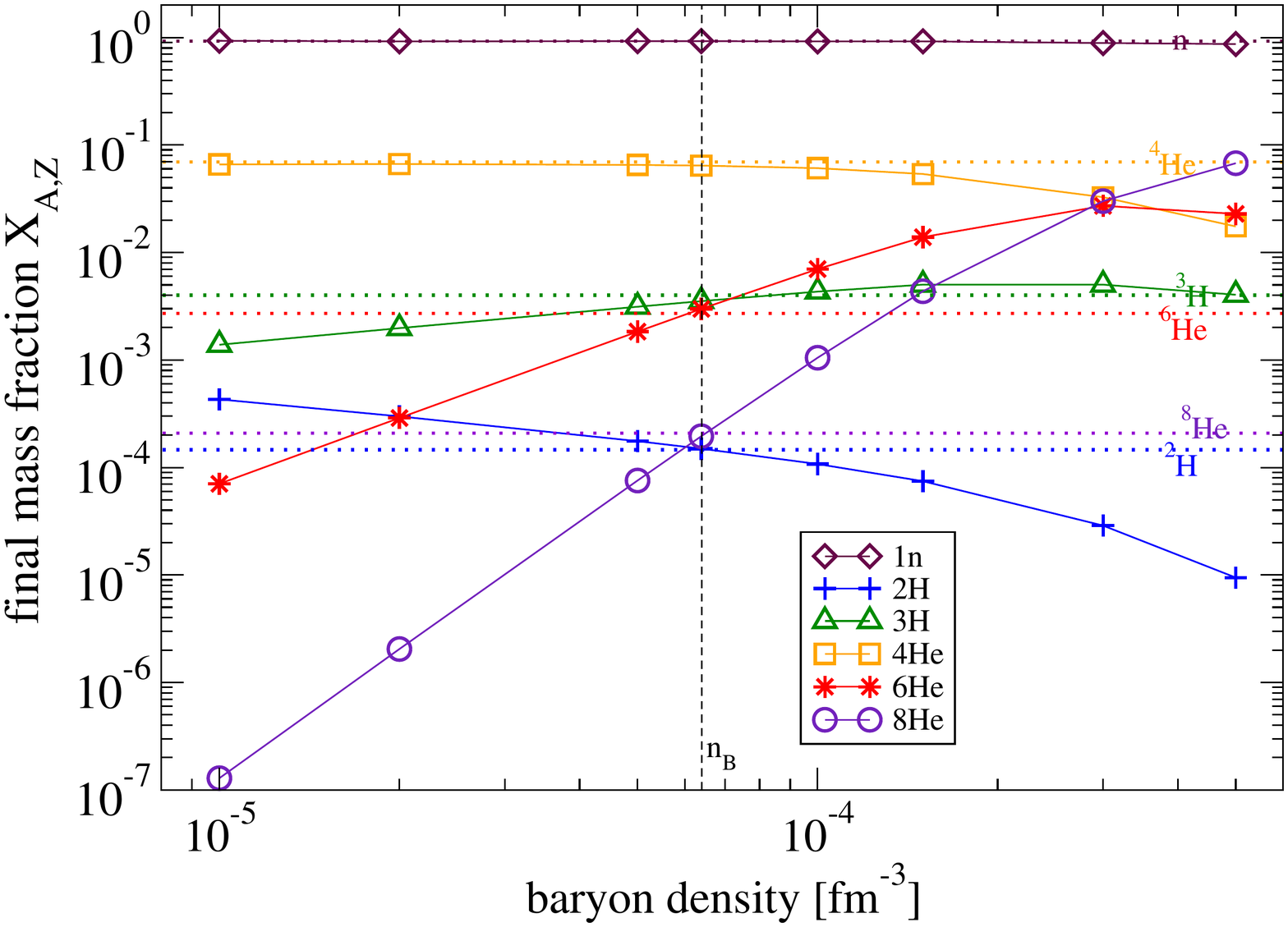}
\caption{
 Calculated light isotope yields (resonance-gas approximation) as a function 
               of baryon density $n_B$ (fixed temperature $T= 1.289$ MeV and proton fraction $Y_p = 0.0349$)
               for $Z=1,2$ isotopes (represented by symbols) are compared with the observed
               experimental yields represented by horizontal dotted lines.
               Optimum agreement based on a fit metric proposed by J. Lestone (\ref{metric}) is found 
               at a density of $n_B=6.4\times 10^{-5}$ fm$^{-3}$ (dashed line). See Figure 2.}
\label{fig:fraction}
\end{figure}

A main result is the low value of the density $n_B=6.4 \times 10^{-5}$/fm$^3$ in resonance-gas approximation. We performed the fit considering the isotopes with $Z \le 2$. Larger LCP ("metals") will be included below in the following section \ref{sec:4}. 
To show how sharp the fit value for the density is, we performed calculations for the composition, mass fraction $\tilde X_{A,Z}=AY_{A,Z}/\sum_{A',Z'\le2}A'Y_{A',Z'}$,
 for fixed values of $\lambda_T$ and proton fraction  $\tilde Y_p=\sum_{A,Z\le2}ZY_{A,Z}/\sum_{A,Z\le2}AY_{A,Z}$ but variable density
$\tilde n_B=\sum_{A,Z \le 2} A Y_{A,Z}$,  divided by the volume, see Fig. \ref{fig:fraction}, where we compare the calculated values (final,res.gas) with the observed values (obs). 
(Note that the densities and proton fractions are calculated in the present section from the yields of $n$, H, and He isotopes, dropping the contributions of the metals.)

As indicated above, all calculations presented in this section were performed for the ideal resonance gas approximation, the simplest of the three cases considered. The same calculations can be performed for the other
approximations of the intrinsic partition function as shown below. 
With respect to the results discussed in the present section, no significant changes in the Lagrange parameters will be observed. However the successive approaches do reveal evidence for continuum and medium effects that even at this very low density, particularly for weakly bound and/or very neutron rich isotopes.

\begin{figure}[h]
\includegraphics[width=280pt,angle=0]{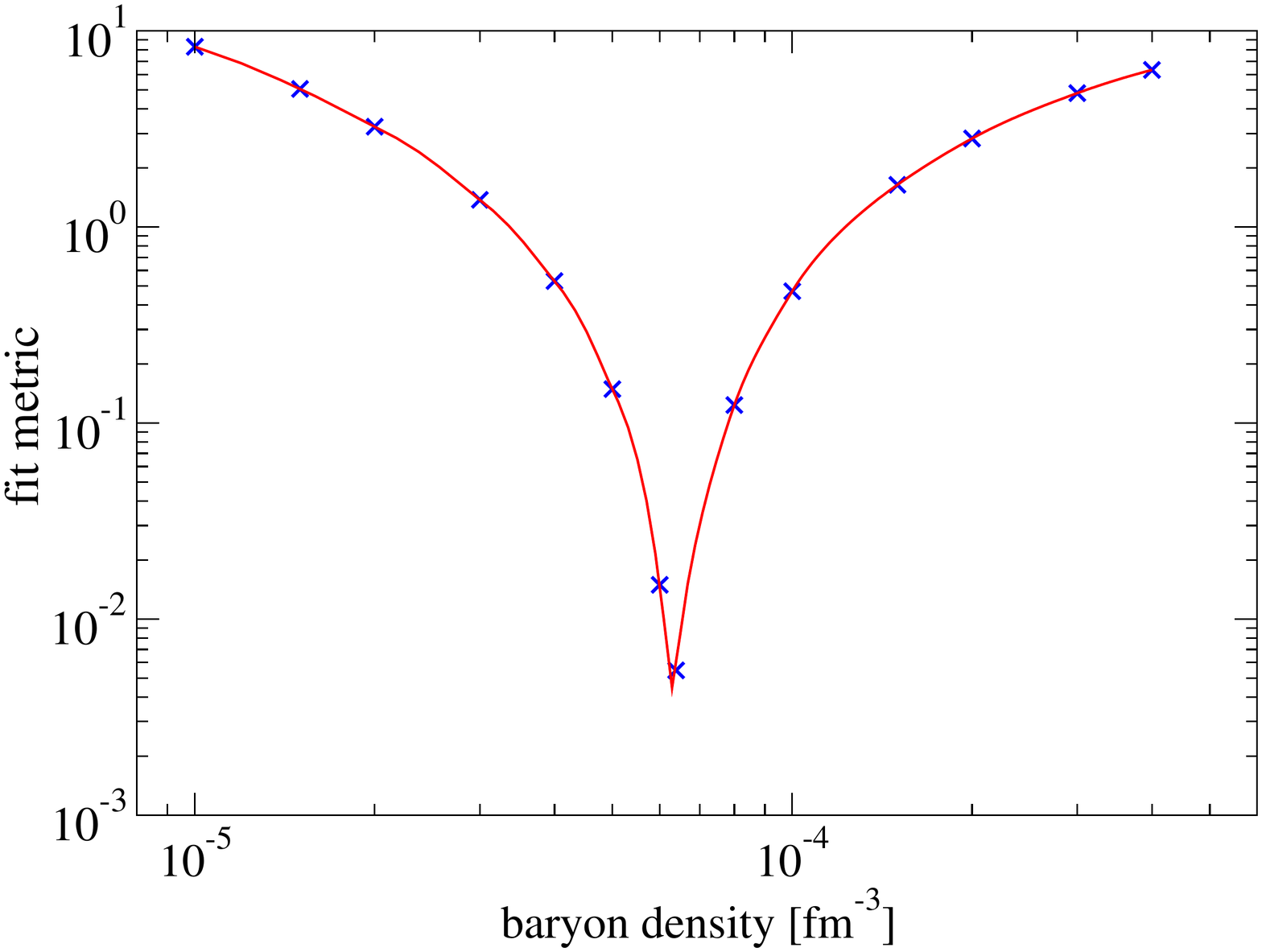}
\caption{
 Fit metric (\ref{metric}) vs. baryon density $n_B$ for $T= 1.289$ MeV and $Y_p = 0.0349$. }
\label{fig:Lestone}
\end{figure}

The quality of the fit is expressed by the Lestone fit metric (\ref{metric}), see Fig.  \ref{fig:Lestone}. 

\section{Inclusion of heavier  isotopes}
\label{sec:4}

We now consider the yields of isotopes with $2 < Z\le 6$. For these elements, accurate values for observed yields of ternary fission are available for of $^{241}$Pu($n_{\rm th}$,f) \cite{Koes1,Koes2}. 
For heavier elements $Z > 6$ calculations can be performed, but observed data become incomplete and less accurate.

For this purpose, we again calculate the relevant, primary distribution within a quantum statistical approach using the several different approximations proposed.
In a first approximation of the ideal resonance gas, for each $A,Z$ not only the ground state, but also all excited states are considered 
which will give the intrinsic partition function. 
We take the excited states from the data tables \cite{Nudat} together with their degeneracy,
see the appendix, Tabs. \ref{tab:He} - \ref{tab:C}. 
The possible decay to other isotopes is also indicated. For this, the threshold energy
$E^{\rm thresh}_{AZ}$ is given where the decay channels open, in general the separation energy $S_n$ for neutrons, but also other possible 
decay channels such as for $^7$Li to $^4$He + $^3$H or $^8$Be to 2$\times ^4$He. In this resonance gas approximation, all known excited states are considered, and states above the threshold energy are assumed to decay and to feed other isotopes in the final distribution.

The second approximation takes the states in the continuum more accurately into account. The virial expansion implements continuum correlations in a systematic way. This virial approximation is also known from the nuclear matter equation of state  \cite{Ropke82,SRS,HS}.

In the third approximation, 
in-medium modifications are taken into account. These are the shifts of binding energies owing to Pauli blocking and possible dissolution of bound states if they are shifted to the continuum \cite{R09,Typel10,R20}.

\subsection{Excited states and resonance gas approximation}

 \begin{table*}
\begin{center}
\hspace{0.5cm}
 \begin{tabular}{|c|c|c|c|c|c|c|c|c|c|}
\hline
isotope& $A$  &  $Z$ &  $\frac{B_{A,Z}}{A} $&  $ g_{A,Z} $  & $E^{\rm thresh}_{AZ}$ & $Y^{\rm obs}_{A,Z}$ &$Y^{\rm obs}_{A,Z}$/$Y^{\rm final, res.gas}_{A,Z}$&$Y^{\rm obs}_{A,Z}$/$Y^{\rm final, vir}_{A,Z}$&$Y^{\rm obs}_{A,Z}$/$Y^{\rm final,medium}_{A,Z}$\\
\hline
1n	&1&0	&0	& 2 				&-		   			& 0.107   			& 0.9742 			& 0.9555   &0.9551 \\
1H	&1& 1 	&0	& 2				&-                & -         			& - 					& -				&-\\
2H 	&2& 1 	& 1.112 & 3 		&2.224			&8.463E-6 		& 0.952 			&0.92			&0.917 \\
3H	&3& 1 	& 2.827 & 2 		&6.257			&1.584E-4   	&1.133 			&1.179			& 1.223 \\
\hline
4He	&4& 2 	& 7.073 & 1 		& 20.577	  	&2.015E-3   	&1.055 			&1.049 		& 1.042 \\
6He	&6& 2 	& 4.878 & 1 		& 0.975		&5.239E-5   	& 0.8856			& 0.8753 		& 0.8903 \\
8He	&8& 2 	& 3.925 & 1 		& 2.125		&3.022E-6   	&1.043  			&1.04 			& 1.03\\
\hline
7Li	&7 & 3 	& 5.606 & 4 	& 2.461		&1.35E-6 		&0.5305 			&0.4964 		& 0.4523  \\
8Li	&8 & 3 	& 5.160 & 5 	& 2.038     	&8.463E-7   	&0.6235 			&0.5669 		& 0.5211\\
9Li	&9 & 3 	& 5.038 & 4 	& 4.062     	&1.672E-6 		&0.7164 			&0.6363 		& 0.5814\\
11Li	&11& 3	& 4.155 & 4 	& 0.396     	&9.068E-10 	&0.03061 		&0.02936 	&  0.2992 \\
\hline
9Be	&9  & 4 	& 6.462 & 4 	& 1.558  		&8.866E-7    	&0.5488 			& 0.4908 		& 0.3191\\
10Be&10 & 4 	& 6.497 & 1 	& 6.497  		&9.269E-6   	&0.6149 			&0.5343 		& 0.5285\\
11Be&11 & 4	& 5.953 & 2 	& 0.502  		&1.189E-6  		&0.2756 			&0.2584  		& 0.4202\\
12Be&12 & 4	& 5.721 & 1 	& 3.17  		&5.642E-7   	&0.1513 			&0.1362 		& 0.1385  \\
14Be&14 & 4	& 4.994 & 1 	& 1.264  		&5.441E-10  	&0.01321 		&0.01326 	& 0.0121\\
 \hline
 11B	&11 & 5 	& 6.928 & 4  	& 8.674		&3.224E-7   	&0.3632 			& 0.3015 		& 0.2257 \\
 12B	&12 & 5 	& 6.631 & 3  	& 3.369  		&2.015E-7   	&0.1098 			&0.09418 	& 0.08358 \\
 14B	&14 & 5 	& 6.102 & 5   	& 0.97 			&2.62E-8  		&0.01742 		&0.01552 	& 0.01793 \\
 15B	&15 & 5 	& 5.880 & 4   	& 2.78 			&9.269E-9   	&0.01205 		& 0.01005 	& 0.008624 \\
 \hline
14C	&14 & 6 	& 7.520 & 1 	& 8.176		&2.539E-6  		&0.0495 			& 0.04598  & 0.02229\\
 15C	&15 & 6 	& 7.100 & 2 	& 1.218		&8.665E-7  		&0.01546 		& 0.01353 	& 0.01464\\
 16C	&16 & 6 	& 6.922 & 1 	& 4.25			&1.008E-6  		&0.01028 		& 0.009781 & 0.007721 \\
  17C&17 & 6 	& 6.558 & 4 	& 0.734 		&1.29E-7  		&0.002748 		& 0.00241  	& 0.005271\\
 18C	&18 & 6 	& 6.426 & 1 	& 4.18  		&5.642E-8   	&0.001276 		& 0.001078 & 0.0006891\\
 19C	&19 & 6 	& 6.118 & 2 	& 0.58 			&5.038E-10 	&0.00003031 & 0.00002647 & 0.0002309\\
 20C	&20 & 6 	& 5.961 & 1 	& 2.98			&7.254E-10  	&0.0001937 	& 0.0001578 & 0.0001285\\
\hline
\hline
$\lambda_T$& - & - & - & - & - & - &1.2897  &1.2913 & 1.2924  \\
$\lambda_n$& - & - & - & - & - & - &-3.1486 &-3.1425 & -3.087\\
$\lambda_p$& - & - & - & - & - & - &-16.273 &-16.236 &-16.189  \\
\hline
volume  &  - & - & - & - & - & - &1859.4	&1876.4 &1796.3  \\
$n_B$   & - & - & - & - & - & - &0.000064 	&0.0000645 & 0.00006741 \\
$Y_p$  & - & - & - & - & - & - &0.03486 	&0.03486 &0.03479  \\
 \hline
fit metric & - & - & - & - & - & - &0.005485 &0.009384 & 0.009574 \\
\hline
\end{tabular}
\caption{Observed yields per fission of ternary fission of $^{241}$Pu($n_{\rm th}$,f)  are compared to a final state distribution, 
obtained from feeding of a relevant (primary) distribution. The primary distribution was calculated in the ideal resonance gas approximation (res.gas), 
as well as for the virial form (vir) and including medium effects (medium).  Lagrange parameters  and other properties are given at the bottom of the Table, the fit metric refers to all isotopes with $Z \le 2$. Units: MeV, fm.} 
\label{Pu241Vir}
\end{center}
\end{table*}

 \begin{figure}[h]
\includegraphics[width=0.5\textwidth]{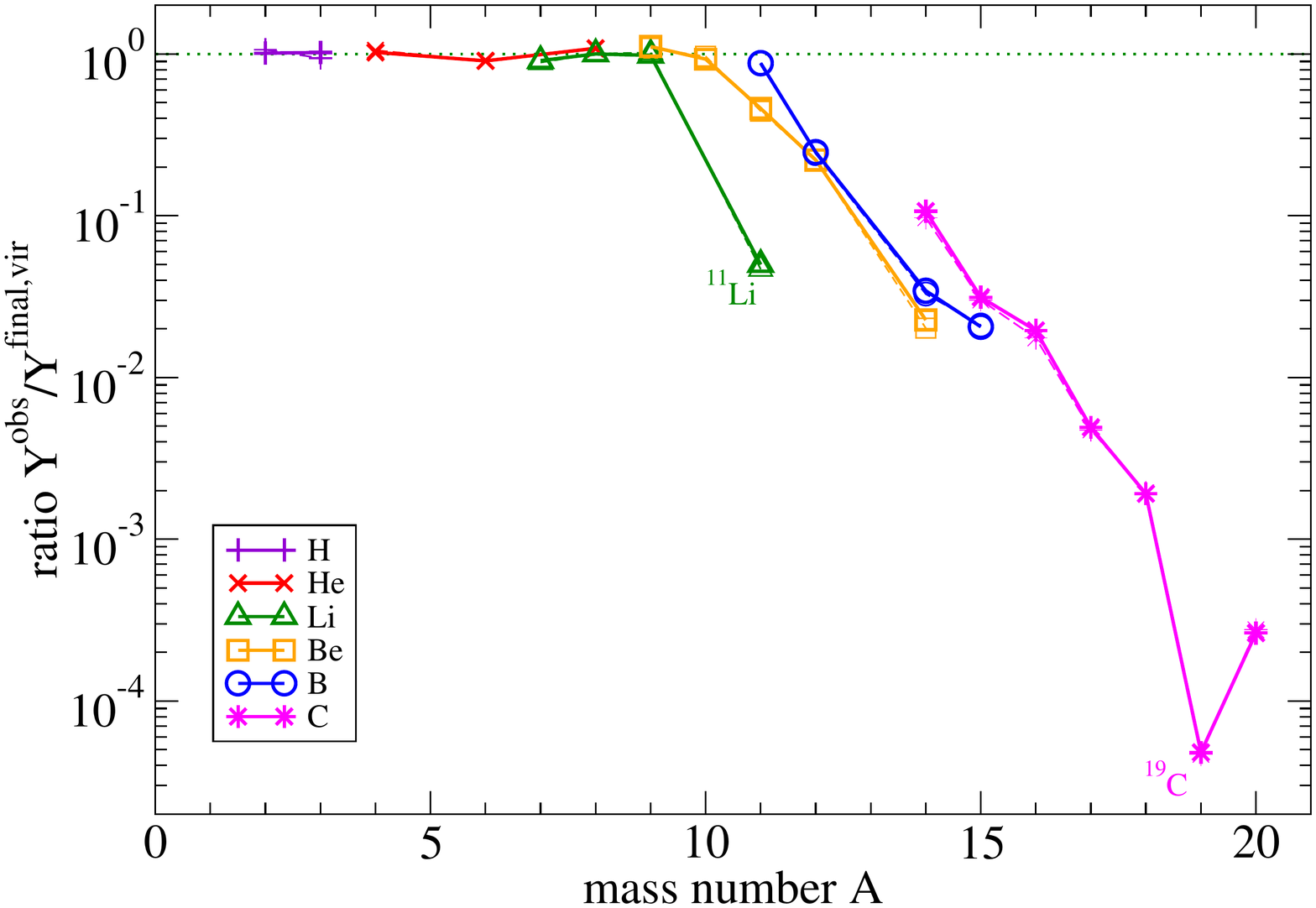}
\caption{
Ternary fission of $^{241}$Pu($n_{\rm th}$,f): Ratios $Y^{\rm obs}_{A,Z}$/Y$^{\rm final, res.gas}_{A,Z}$ (bold symbols) and $Y^{\rm obs}_{A,Z}$/Y$^{\rm final, vir}_{A,Z}$ (full lines) as function of the mass number $A$. Isotopes with $Z \le 6$ are shown. Data  from Tab.  \ref{Pu241Vir}. }
\label{fig:excited}
\end{figure}

  Data for excited states are shown in the appendix, Tabs. \ref{tab:He} to \ref{tab:C}. In addition, the threshold energy $E^{\rm thresh}_{AZ}$ is shown for the continuum of scattering states. 
 As indicated above, this is in general the neutron separation energy $S_n$, but other decay channels (e.g. triton, $\alpha$) are also possible. 
 For each isotope of the primary distribution, different final states are possible if excited, unstable states are considered. 
 A branching ratio would indicate the ratio to final state transitions during the expansion after freeze out.
 
 We use the simple approximation that all excited states below the continuum edge decay to the ground state of the same isotope.
 States above the continuum edge decay to other final isotopes. 
 The same happens also with the unbound nuclei like $^4$H, $^5$He. These states which feed down to other isotopes are marked with an asterisk *, and the process is indicated
 ($\to ^3$H for $^4$H, etc.). Weakly bound states within 1 MeV below the threshold energy are marked with "$^0$". 
 In the present approximation of the ideal resonance gas, we assume that these states contribute to the ground state final yield of the same isotope after de-excitation. 
 The corresponding factors $R_{A,Z}^{\rm res.gas}(\lambda_T)$ which are related to the intrinsic partition function are calculated for $\lambda_T=1.29$ MeV in  Tabs. \ref{tab:He} to \ref{tab:C}. 
 These partition function multipliers are also shown in Tabs. \ref{241Pujun1}, \ref{241Pujun2}, together with the relevant, primary yields $Y^{\rm rel, res.gas}_{A,Z}$, and the final yields $Y^{\rm final, res.gas}_{A,Z}$.
 
To compare with the observed yields, results for the ratio $Y^{\rm obs}_{A,Z}$/$Y^{\rm final, res.gas}_{A,Z}$ for this resonance gas approximation are shown in Tab. \ref{Pu241Vir} as well as in Fig. \ref{fig:excited}. 
The global behavior can be described as follows: Up to $A = 10$ the observed yields are rather well reproduced by the grand canonical equilibrium calculations, for larger $A$ we see a strong suppression, as has already been discussed in \cite{Sara, rnp20,rnp21}. Individual isotopes show deviations from the global behavior which may be caused by the approximations in calculating the final distribution, i.e., neglecting interaction effects. This clearly bears further investigation.

\subsection{Virial expansion and continuum correlations}

As previously indicated we have improved the equilibrium calculations in two successive steps. First we consider the virial expansion which accounts for continuum contributions.
In the subsequent section, in-medium effects are also taken into account. Results for the virial approximation are shown in Tab. \ref{Pu241Vir}
and Fig. \ref{fig:excited}.

The cluster-virial expansion considers continuum correlations of two constituents of nuclear matter in terms of the scattering phase shifts, 
as obtained in correspondence to the Beth-Uhlenbeck formula. For instance, $^4$H is considered as resonance in the $t-n$ channel,
where scattering phase shifts have been measured, see \cite{R20,HS}. For details see the appendix \ref{sec:vir}. 
Data for $R_{A,Z}^{\rm vir}(T)$ are shown in the appendix, Tabs. \ref{tab:He} to \ref{tab:C}. 
The resulting ratios $Y^{\rm obs}_{A,Z}$/$Y^{\rm final, vir}_{A,Z}$ are also shown in Tab. \ref{Pu241Vir}.

We see that the changes are small in general, so that the influence of the continuum correlations on the final yield distribution is not essential.
This may be attributed to the low temperature so that the contribution of scattering states is small for binding energies larger than $T$.
At higher $T$, the contribution of scattering states would  become more important. The most important changes are seen for $^4$H so that the feed-down to the final $^3$H is reduced. Some He isotopes are also 
reduced. The optimum fit of Lagrange parameters is also slightly changed. 
However, for several of the isotopes considered, strong deviations from the global behavior remain.
 
  As discussed in the case of $^6$He \cite{rnp21}, weakly bound states are more sensitive to medium effects. 
The threshold energy of $^{11}$Li is low, and the yield is strongly overestimated within the ideal gas and virial approaches. 
Obviously, the measured yield of $^{11}$Li is not well described in the approximations where the interaction between the components 
 is neglected.

Low yields of other 'exotic' nuclei ($^{19}$C etc.) are also known, and are observed in ternary fission of other actinides (Am, Cf, etc.).
For the carbon isotopes, the spectrum of excited states is rather  complex and feeds different final states. In particular, isotopes with excited states above the 
neutron separation energy feed the yield of isotopes with $A-1$. Calculations of the yields, see Figs. 3 and 4 in \cite{Koestera}, cannot explain the low yield for $^{19}$C. 
As seen there, the authors offer no explanation for the low yield of these exotic neutron rich, halo-like nuclei.


\subsection{In-medium effects}
 \label{inmedium}

 \begin{figure}[h]
\includegraphics[width=0.5\textwidth]{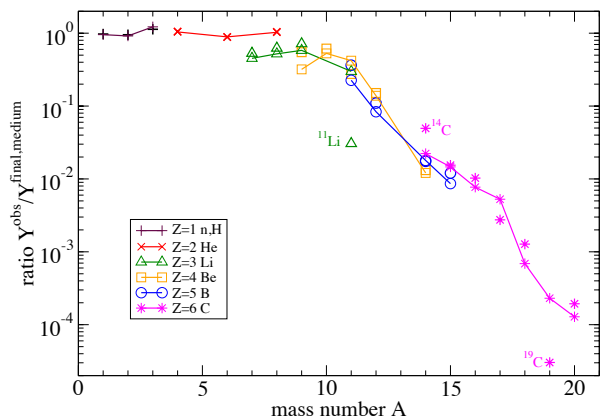}
\caption{
Ternary fission of $^{241}$Pu($n_{\rm th}$,f): Ratios $Y^{\rm obs}_{A,Z}$/Y$^{\rm final, res.gas}_{A,Z}$ (bold symbols) and $Y^{\rm obs}_{A,Z}$/Y$^{\rm final,medium}_{A,Z}$ (full lines) as function of the mass number $A$. Isotopes with $Z \le 6$ are shown. Data from Tabs. \ref{Pu241Vir}, \ref{241PuG}, \ref{241PuH}.}
\label{fig:medium}
\end{figure}

   We now consider two medium effects which might modify the yield distributions: Self-energy shifts and Pauli blocking. Self-energy shifts act on all nucleons and can be accounted for by a renormalization of the chemical potentials $\tilde \lambda_n, \tilde \lambda_p$ if they are not momentum dependent. Ratios of yields are not influenced by these shifts. In contrast, Pauli blocking acts individually for each isotope and leads to strong deviations of the yield distribution.
  
  Self-energy of nucleons has been treated, for instance, within the relativistic mean-field approximation. The well-established parametrization of DD2-RMF \cite{Typel} gives for the parameter values
  $T=1.29$ MeV, $n_B=0.000067$ fm$^{-3}$, $Y_p=0.035$ the self-energy shifts $\Delta_n^{\rm SE}=-0.03934$ MeV,  $\Delta_p^{\rm SE}=-0.09936$ MeV. 
  For a cluster $\{A,Z\}$ the self-energy shift is $\Delta E^{\rm SE}_{A,Z}=(A-Z)\Delta_n^{\rm SE}+Z\Delta_p^{\rm SE}$. We then have $\tilde \lambda_n=\lambda_n-\Delta_n^{\rm SE}, \tilde \lambda_p=\lambda_p-\Delta_p^{\rm SE}$.

 In all our earlier fits \cite{rnp21}, the yields of $^6$He are overestimated, whereas the yields of $^8$He are underestimated. One possible explanation is that $^6$He is only weakly bound ($E^{\rm thresh}=0.975$ MeV) compared to $^8$He ($E^{\rm thresh}=2.125$ MeV). In the dense medium, binding energies are shifted, and weakly bound states are more affected than strongly bound states.
 
 More visible in Fig. \ref{fig:excited}  is the strong suppression of weakly bound isotopes such as $^{11}$Li and $^{19}$C. 
 These isotopes are near to the continuum edge so that the shift owing to Pauli blocking may lead to dissolution (the Mott effect \cite{Ropke82}). 
 This suggests the use of yields of weakly bound states as test probes to infer the free neutron density. The observation of such  Mott effects would provide us with an independent test of the free neutron density at scission. 
 
To quantify this effect, 
we consider the bound state contribution to the intrinsic partition function which contains the factor $e^{-E^{0}_{A,Z}/T}-1$ \cite{rnp20}.  If the bound state energy is shifted owing to in-medium effects,
 this factor  is also changed and goes to zero at the Mott density where bound states disappear.
 
The Pauli blocking shift of bound states in dense matter and the Mott effect has previously been considered in detail for individual isotopes \cite{Ropke82,R09,R20}.  We will give here only a general estimate applicable to bound states of all nuclei. 
 The shift of bound state energies $E_{A,Z}(T,n_B,Y_p)-E^{0}_{A,Z} $ in a dense medium (Pauli blocking) has been estimated in \cite{R20}, Eq. (22), for nucleons in the $1s$ and $2p$ orbits as
 \begin{equation}
\Delta E_{A,Z}^{\rm Pauli} = 1064 (A-Z) n_B  e^{-0.0513/T} {\rm MeV\, fm}^3
\end{equation}
 where we assume that the proton contribution is negligible because of its very low density. With the baryon density $n_B=6.7 \times 10^{-5}$ fm$^{-3}$ (which should be determined self-consistently), the shifts of the binding energy of the different isotopes are given in Tabs. \ref{241PuG} and \ref{241PuH}.
 These shifts may become important for weakly bound states. For instance, for $^{11}$Li  the shift is larger than the threshold energy so that the bound state is dissolved. This is also reflected in the low observed yield of this isotope. 
 In a more detailed calculation  \cite{R11} not considered here, the Pauli blocking shift depends on the center-of-mass momentum $P$ of the cluster.
 Because the shift becomes smaller for larger total momentum $P$, and continuum correlations may be present, 
 a small yield of this isotope remains. The same happens also with $^{19}$C where the shift is also larger than the threshold energy so that the Mott effect leads to a significant reduction of the observed yields. This is also seen in Fig. \ref{fig:medium}. The yields of other isotopes with a low threshold energy ($E^{\rm thresh}_{AZ} < 1$ MeV) will be significantly influenced too. The excited states $^{11}$Be$^0$, $^{14}$B$^0$, $^{15}$C$^0 - ^{17}$C$^0$ become dissolved because of Pauli blocking. The corresponding calculations are seen in Tabs.
 \ref{241PuG},  \ref{241PuH}. 
 
 As shown from Tab. \ref{Pu241Vir} and Fig. \ref{fig:medium}, the observed yields are, in fact, sensitive to in-medium effects. Suppression effects owing to Pauli blocking seem to be visible. 
 This allows an independent second determination of the baryon density, and the densities at which such isotopes which are dissolved are consistent with the inferred value $n_B=6.7 \times 10^{-5}$ fm$^{-3}$.
 Note that a fully ab initio calculation of the Pauli blocking effects is very involved, and we gave here only exploratory calculations to illustrate the effect of suppression.

\section{Generalized relativistic mean-field approach}

As discussed above, important in medium effects are self-energy shifts and Pauli blocking. If the self-energy shifts are not dependent on the nucleon momentum, they can be absorbed into the chemical potential thus creating an effective chemical potential.
Different approximations my be used to describe the in-medium self-energy shifts, for instance the Skyrme or the relativistic mean-field approaches. 

In Ref.~\cite{Pais1}, Pais {\it et al.} reported a generalized  relativistic mean-field (RMF) approach formulated for the study of in-medium modifications on light cluster properties. Explicit binding energy shifts and a modification of the scalar cluster-meson coupling were introduced in order to take these medium effects into account. The interactions of the clusters $i=\,^2$H, $^3$H, $^3$He, $^4$He with the surrounding medium are described with a phenomenological modification, $x_s$, of the coupling constant to the $\sigma$ meson, $g_{\sigma_i}= x_s A_i g_\sigma$, where $g_\sigma$ is the nucleon scalar coupling, and $A_i$ the number of nucleons in cluster $i$. Using the FSU Gold EoS \cite{FSU}, 
 and requiring that the cluster fractions exhibit the correct behavior in the low-density virial limit \cite{virial,Horowitz06}, they obtained a universal scalar cluster-meson coupling fraction, $x_{\sigma_i} = 0.85 \pm 0.05$, which could reproduce both this limit and the equilibrium constants extracted from reaction ion data \cite{Qin12} reasonably well. Later this work was generalized to include clusters up to $A=12$ \cite{Pais3}.
 
 In a more recent work \cite{PaisExp}, 
 Pais {\it et al.} compared results of an 
 analysis, where in-medium effects are included, to experimental equilibrium constants measured in intermediate energy Xe + Sn collisions. This comparison required a higher scalar cluster-meson coupling constant $x_{\sigma_i} = 0.92 \pm 0.02$. With this higher assumed value of the coupling constant, the in-medium effects are reduced, and the clusters melt at larger densities.

\begin{figure}[h]
\begin{tabular}{c}
\includegraphics[width=0.85\linewidth]{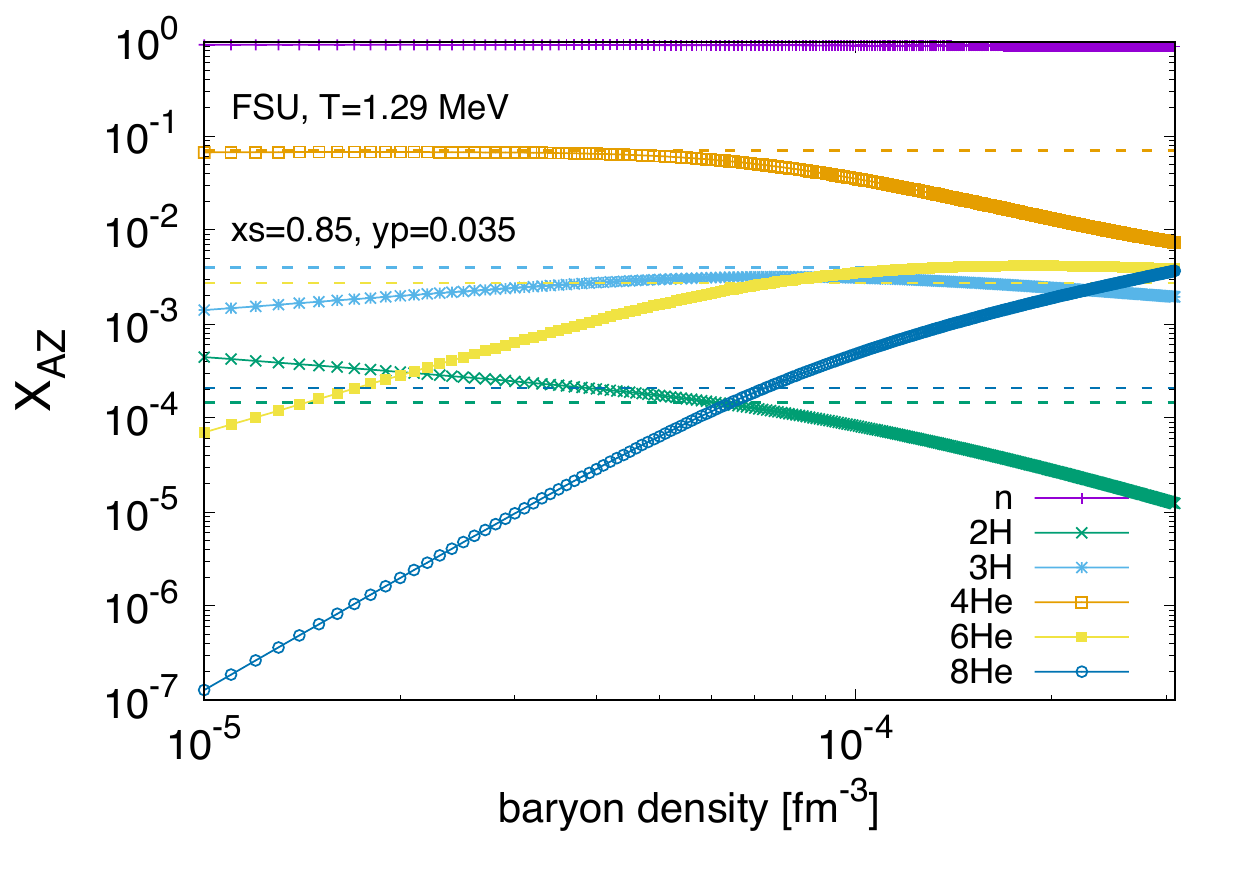} \\
\includegraphics[width=0.85\linewidth]{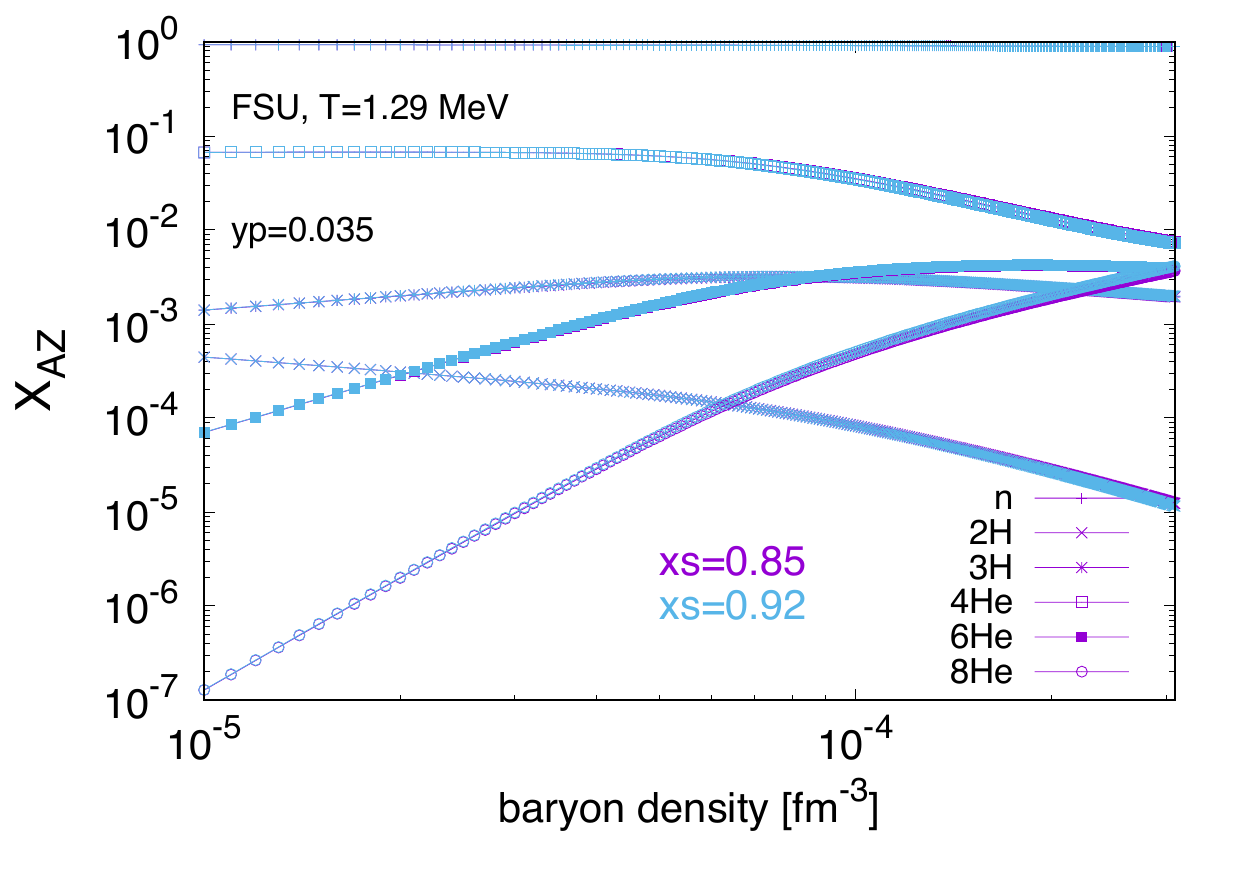}
\end{tabular}
\caption{(Top) Mass fractions as a function of the density for the $Z=1,2$ isotopes and the free neutron gas in a RMF calculation with in-medium effects ($x_s=0.85$) for the FSU model with a fixed temperature of $T= 1.29$ MeV, and a proton fraction of $Y_p = 0.035$. The observed experimental yields are represented by horizontal dotted lines.  (Bottom) Mass fractions of the light particles as a function of the density in a RMF calculation employing different values of the scalar-cluster-meson coupling  constant. } 
\label{fig:rmf-mass}
\end{figure}

In the top panel of Fig.~\ref{fig:rmf-mass}, we show the mass fractions of the clusters with $Z=1,2$ and also the fraction of the free neutron gas, calculated using the RMF formalism with $x_s=0.85$ , as a function of the density, for the temperature and proton fraction found in the QS approach with in-medium effects described in the previous subsection. In the bottom panel, the mass fractions of light particles are plotted against the density, for different values of the scalar cluster-meson coupling. We observe that the abundances shown in the top panel are quite similar to the ones in the resonance gas approximation (see Fig. 1). Also, from the bottom panel, we see that the medium effects are almost negligible because the densities are very small. As it was shown in Tab. III, the effect of the medium is more relevant when considering clusters with higher $Z$.
 
\begin{figure}[h]
\includegraphics[width=0.5\textwidth]{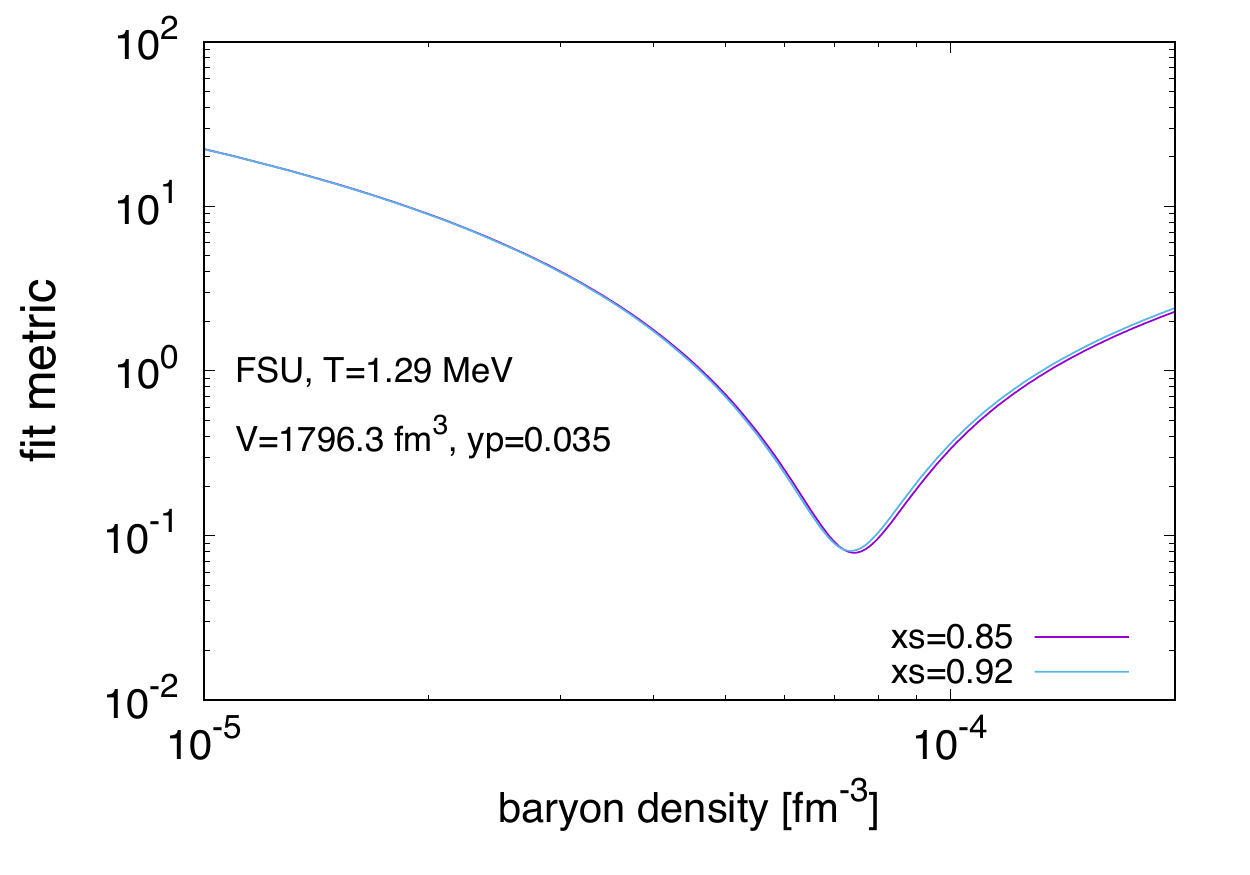}
\caption{
 Fit metric (\ref{metric}) vs. baryon density for $T= 1.29$ MeV and $Y_p = 0.035$, using different values of the scalar cluster-meson coupling. }
\label{fig:fitmetric-rmf}
\end{figure}

In Fig.~\ref{fig:fitmetric-rmf}, we present the calculated fit metrics derived with different values of the scalar cluster-meson coupling constant, and we observe a minimum at about the same density as in the QS approach, $n_B \sim 6.7 \times 10^{-5}$ fm$^{-3}$.

\section{Conclusions}

We have considered the modeling of ternary fission yields within a systematic quantum statistical approach and a generalized relativistic mean field approach. We consider our quantum statistical approach as a first step in a strict non-equilibrium approach.
As an example, we have analyzed the observed yields for $^{241}$Pu($n_{\rm th}$,f).

We assume that the light charged fragments 
are emitted from the neck region at scission. We have first considered partial chemical equilibrium for clusters $Z \le 2$.
During the further expansion, feed-down processes were considered.
This leads to the final yields which are compared to the observed yields. Higher mass clusters are suppressed by nucleation kinetics and/or size effects.

We are interested in an optimal description of the primary equilibrium distribution at scission.
Within a quantum statistical approach, excited bound states and continuum correlations are taken into account.
In particular, we have searched for in-medium effects going beyond a simple statistical model of non-interacting 
components (ideal gas). These in-medium effects are determined by the thermodynamic parameters of matter in the scission region.
For all nuclei $Z \le 2$, the Lagrange parameter values are: $\lambda_T=1.29,\,\,\lambda_n=-3.09 ,\,\,\lambda_p=-16.19$ MeV,  volume 1796 fm$^3$, neutron density $n_n=6.7 \times 10^{-5}$  fm$^{-3}$.
We showed that for those parameter values, density effects are expected and are manifested by the experimental data.

In particular we find evidence for the Mott effect for the weakly bound states $^{11}$Li and $^{19}$C, and also the suppression of the yields for weakly bound nuclei $^6$He, $^{11}$Be, $^{14}$B, and  $^{15}$C $- ^{17}$C (excited states). This shows the consistency of our approach: the nucleon density  $6.7 \times 10^{-5}$ fm$^{-3}$ inferred from composition is also seen in the in-medium shifts, the weakly bound clusters serving as  test probes. We identify the strong suppression of $^{11}$Li and $^{19}$C as a signature of the Mott effect and find that other isotopes are not dissolved so that we find an estimate for the Pauli blocking effect and the corresponding interval of density.  

We also considered a generalized relativistic mean-field approach to study the in-medium modifications on light cluster properties, introducing a modification on the scalar cluster-meson coupling. 
At the low nucleon densities considered here, no significant changes in the yield distribution of isotopes and the nuclear-matter parameter values have been observed.
This underlines the validity of the parameter values for temperature, nucleon density and proton fraction, presented in this article, to describe nuclear matter in the neck region at scission.

The improved description of nuclear matter at scission conditions is of interest for the properties of neutron rich astrophysical systems \cite{Bezn20,Fischer,Furusawa22,Gulm,Lim,Du,Sedr}, for instance the equation of state or the neutrino opacity, at similar thermodynamic parameter values.

\section*{ACKNOWLEDGMENTS}
This work was supported by the United States  Department of Energy under Grant No. DE-FG03-93ER40773. It was partly supported by the FCT (Portugal) Project No. UID/FIS/04564/2020. H.P. acknowledges the grant CEECIND/03092/2017 (FCT, Portugal). G.R. acknowledges support by the German Research Foundation (DFG), Grant \# RO905/38-1.

\appendix

\section{Excited state multiplier and intrinsic partition function}

The contribution of excited states to the intrinsic partition function of a particular channel $\{A,Z\}$ which after expansion (feed-down) decays to a different final isotope, observed in experiment (branching ratio), is denoted by *.
We separate also the weakly bound states which are within 1 MeV below the edge of continuum. We denote it by "$^0$" below. 
Sometimes we separate the ground state so that "$^0$" appears twice.

The threshold energy $E^{\rm thresh}_{AZ}$ is generally given 
by the neutron separation energy, 
but lower threshold energies may appear for other decompositions, for instance $^6$Li $\to \alpha + d$ and $^7$Li $\to \alpha + t$, see 
\cite{Jesinger05}. This defines feed down channels and the corresponding subdivisions of the intrinsic partition function.

The corresponding partition function multipliers
 $R^{\rm res.gas}_{AZ}(\lambda_T)$ and $R^{\rm vir}_{AZ}(\lambda_T)$ are calculated for $\lambda_T=1.29$ MeV. The resulting yields $Y^{\rm rel,vir}_{A,Z}$ 
are given in Tabs. \ref{tab:He} -  \ref{tab:C} for the elements up to carbon, see also the Supplemental material of Ref. \cite{rnp21}.

\subsection{$Z \le 2$: n, H, He}
\label{App:1}

\begin{table*}[h]
\begin{center}
 \begin{tabular}{|c|c|c|c|c|c|c|c|c|c|c|}
\hline
isotope&feeddown& $A$  &  $Z$ &  $B_{A,Z}/A $&  $ g_{A,Z} $  & $E^{\rm thresh}_{AZ}$ & $E_i\,\, [g_i]$ & $R_{A,Z}^{\rm res.gas}( 1.29)$& $R_{A,Z}^{\rm vir}( 1.29)$ & $R_{A,Z}^{\rm medium}( 1.29)$  \\
\hline
$^1$n	&-			&1 &0& 0		& 2 	&-			&-																	&1 		&1			&1		\\
$^1$H	&-			&1 &1& 0		& 2	&-              		&-														&1       &1			&1	 	\\
$^2$H 	&-			&2 &1& 1.112 	& 3 	&2.224		&-													&1	 	&0.974		&0.9306	\\
$^3$H	&-			&3 &1& 2.827 	& 2 	&6.257		&-														&1	 	&0.999		&0.9138	\\
$^4$H*	&$\to ^3$H	&4 &1& 1.720	& 5 	&-1.6		&0.31 [3], 2.08 [1], 2.83 [3]	&1.579	&0.09404	&0.03564	\\
\hline
$^3$He	&-			&3 &2& 2.573	& 2 	& 5.494		&-													& 1	 	&0.9979	&0.9545	\\
$^4$He	&-			&4 &2& 7.073 	& 1 	& 20.577		&20.21 [1]								& 1 		&1			&0.915	\\
$^5$He*	&$\to ^4$He	&5 &2& 5.512	& 4 	&-0.735		&-										& 1 	 	&0.6906	&0.5029	\\
$^6$He$^0$&[$\to ^4$He]	&6 &2& 4.878 	& 1 	& 0.975	&-								& 1 		&0.9343	&0.7723	\\
$^6$He*	&$\to ^4$He	&6 &2& 4.878 	& 1 	& 0.975		&1.797 [5]					&1.242 &0.7919	&0.4617	\\
$^7$He*	&$\to ^6$He	&7 &2& 4.123	& 4	& -0.410		&2.92 [6]							&1.156	&0.9313	&0.6577	\\
$^8$He	&-			&8 &2& 3.925 	& 1 	& 2.125		&-												& 1 		&0.972		&0.7383	\\
$^8$He*	&$\to ^4$He	&8 &2& 3.925 	& 1 	& 2.125		&3.1 [5]						&0.452&0.2334&0.06923	\\
$^9$He*	&$\to ^8$He	&9 &2& 3.349	& 2 	& -1.25		&1.1 [2]								&1.426 &0.364	&0.04882	\\
\hline
 \end{tabular}
\caption{Data of  nucleons/nuclei $Z \le 2$ [units: MeV, fm]. Feed-down to other final isotopes indicated by $\to ^A$Z. Mass number  $A$, charge number $Z$. 
Ground state binding energy $B_{A,Z}$ and degeneracy $g_{A,Z}=2J+1$, continuum threshold energy $E^{\rm thresh}_{AZ}$, excitation energy $E_i$ and degeneracy $g_i$   according \cite{Nudat}.  
Partition function multiplier $R_{A,Z}^{\rm res.gas}(T)$  according (\ref{Rfactor2}). 
The contribution of excited states above the continuum threshold is denoted by "*", the contribution of states within 1 MeV below the threshold is denoted by "$^0$" (weakly bound states).}
\label{tab:He}
\end{center}
\end{table*}

Excited states and intrinsic partition function multipliers $R$ of  H and He isotopes are shown in Tab. \ref{tab:He}. The contribution of excited states which after freeze-out decays to a different channel $\{A, Z\}$ (feed-down) is denoted by $^*$.
At low temperatures considered in this work, the subnuclear excitations of the nucleons $n, p$ are neglected so that $R_{1,0}=R_{1,1}=1$.

\subsection{Lithium}

\begin{table*}[h]
\begin{center}
 \begin{tabular}{|c|c|c|c|c|c|c|c|c|c|c|}
\hline
isotope&feeddown& $A$  &  $Z$ &  $B_{A,Z}/A $&  $ g_{A,Z} $  & $E^{\rm thresh}_{AZ}$ & $E_i\,\, [g_i]$ & $R_{A,Z}^{\rm res.gas}( 1.29)$ & $R_{A,Z}^{\rm vir}( 1.29)$& $R_{A,Z}^{\rm medium}( 1.29)$  \\
\hline
6Li		&-			&6 & 3 	& 5.332	& 3 		&1.475  & - 																						&1 			&0.9544	&0.83	\\
6Li*		&$\to ^3$H	&6 & 3 	& 5.332	& 3 	&1.475 &2.186 [7], 3.563 [1],  4.312 [5], 5.366 [5], 5.65 [3] 	&0.5471	&0.3847	&0.2842	\\
7Li		&-			&7 & 3 	& 5.606	& 4 		& 2.461 & 0.478  [2]																		&1.345 	&1.316		&1.097	\\
8Li		&-			&8 & 3 	& 5.160 	& 5 		& 2.038 & 0.980 [3]																		&1.281  	&1.242		&0.9877	\\
8Li*		&$\to ^7$Li 	&8 & 3 	& 5.160 	& (5) 	& 2.038 & 2.255 [7], 3.210 [3], 5.4 [3] , 6.1 [7]				&0.3151 	&0.2646	&0.1975	\\
9Li		&-			&9 & 3 	& 5.038 	& 4 		& 4.062 & 2.691 [2], (4.301) [?]														&1.062 	&1.055		&0.8069	\\
10Li*		&$\to ^9$Li	&10 &3  	& 4.531	&{\it  (3)} 	& -0.032& -                     											&1			&0.863		&0.5866	\\
11Li$^0$	&[$\to ^9$Li]	&11& 3	& 4.155 	& 4 		& 0.396 & (1.266) [?]      												&1 			&0.9003	&0.6026	\\
12Li*		&$\to ^{11}$Li	&12& 3	& 3.792 	&{\it  (3)}	& -0.201& -      													&1			&0.8422	&0.4697	\\
\hline
 \end{tabular}
\caption{Data of  lithium nuclei [units: MeV, fm]. Mass number  $A$, charge number $Z$. Ground state binding energy $B_{A,Z}$ and degeneracy $g_{A,Z}$, excitation energy $E_i$ and degeneracy $g_i$ as well as continuum threshold energy $E^{\rm thresh}_{AZ}$ according \cite{Nudat}. 
Partition function multiplier $R_{A,Z}^{\rm res.gas}(\lambda_T)$ for $\lambda_T=1.29$ MeV. 
 $^7$Li: $E^{\rm thresh}_{AZ}$  for decay to $^4$He+$^3$H; excited states at 4.630 MeV [8], 6.680 MeV [6] decay this way, but also IT.  
}
\label{tab:Li}
\end{center}
\end{table*}

The yields of Li isotopes are shown in Tab. \ref{tab:Li}. We consider $6 \le A \le 12$ 
The degeneracy is not known for $^{10}$Li and  $^{12}$Li, the value 3 was assumed, but is not of relevance for the discussion here.
The same holds also for some excited states.
The threshold energy $E^{\rm thresh}_{AZ}$ is given in 
general by the neutron separation energy, 
but lower threshold energies appear for $^6$Li $\to \alpha + d$ and $^7$Li $\to \alpha + t$, see 
\cite{Jesinger05}. 
 

\subsection{Beryllium}

\begin{table*}[h]
\begin{center}
\hspace{0.5cm}
 \begin{tabular}{|c|c|c|c|c|c|c|c|c|c|c|}
\hline
isotope&feeddown& $A$  &  $Z$ &  $B_{A,Z}/A $&  $ g_{A,Z} $  & $E^{\rm thresh}_{AZ}$ & $E_i\,\, [g_i]$ & $R_{A,Z}^{\rm res.gas}( 1.29)$& $R_{A,Z}^{\rm vir}( 1.29)$ &$R_{A,Z}^{\rm medium}( 1.29)$  \\
\hline
7Be			&-				&7 &4 	& 5.372	& 4 	& 1.585 	& 0.429  [2]  																	& 1.359	&1.302		&1.1324\\
8Be*			&$\to ^4$He		&8 &4 	& 7.062	& 1 	& -0.088	& 3.03 [5]             												& 1.477  	&1.266		&1.0192	\\
9Be			&-				&9 &4 	& 6.462	& 4 	& 1.558 	& -     																				& 1			&0.9572	&0.7585	\\
9Be*			&$\to ^4$He&9 &4 & 6.462	& (4) & 1.558 	& 1.684 [2], 2.429 [6], 2.78 [2], 3.049 [6] 				&0.5628	&0.4796	&0.3629	\\
10Be			&-				&10&4 	& 6.497 	& 1 	& 6.497 	& 3.368 [5]   															&1.367		&1.366		&1.0462\\
10Be$^0$		&[$\to ^9$Be]		&10&4 	& 6.497 	& (1) & 6.497 	& 5.958 [5], 5.959 [3] 							&0.0789	&0.07181	&0.05348	\\
11Be$^0$		&[$\to ^{10}$Be]	&11& 4	& 5.953 	& 2 	& 0.502 	& -         												& 1			&0.9076	&0.6422		\\
11Be$^0$		&[$\to ^{10}$Be]	&11& 4	& 5.953 	& (2) & 0.502 	& 0.320 [2]         									&0.7803	&0.6894	&0.4802\\
11Be*		&$\to ^{10}$Be	&11&4&5.953& (2) & 0.502& 1.783 [6], 2.654 [4], 3.4 [4], 3.889 [4], 3.955 [4]	&1.343		&0.308		&0.03957\\
12Be			&-				&12&4	& 5.721	& 1 	& 3.170 	& 2.109 [5], 2.251 [1]													& 2.149		&2.122		&1.4798\\
12Be$^0$		&[$\to ^{10}$Be]	&12&4	& 5.721	&(1) & 3.170 	&2.715 [3]												&0.3657	&0.3307	&0.222	\\
13Be*		&$\to ^{12}$Be		&13&4	& 5.241	& 2 	& -0.510	& -																	& 1			&0.7806	&0.3144		\\
14Be			&-				&14&4	& 4.994 	& 1 	& 1.264 	& -   																				& 1			&0.9468	&0.5897	\\
15Be*		&$\to ^{14}$Be		&15&4	& 4.541	& 6 	&-1.800	& -   																	& 1			&0.02241	&0.0005243	\\
\hline
 \end{tabular}
\caption{Data of  beryllium nuclei [units: MeV, fm]. 
Mass number  $A$, charge number $Z$, observed yields  $Y^{\rm obs}_{A,Z}$ \cite{Koes1}. Ground state binding energy $B_{A,Z}$ and degeneracy $g_{A,Z}$, excitation energy $E_i$ and degeneracy $g_i$ as well as continuum threshold energy $E^{\rm thresh}_{AZ}$ according \cite{Nudat}.   Excited state multiplier $R_{A,Z}^{\rm excited}(\lambda_T)$ for $\lambda_T=1.29$ MeV. Weekly bound isotopes $^{10}$Be$^0$, $^{11}$Be$^0$, 
and $^{12}$Be$^0$ are separated, they will merge with the continuum at higher densities.
}
\label{tab:Be}
\end{center}
\end{table*}

\hspace{0.5cm}

Results for Be isotopes are shown in Tab. \ref{tab:Be}. For the relevant, primary distribution we have to consider also the unstable, excited states. Excited states above the continuum edge which emit a neutron or other nuclei are taken separately, their contribution to the intrinsic partition function is denoted by an asterisk. Also excited states little below the continuum edge (1 MeV) may become dissolved in a dense medium if the binding energy is reduced owing to Pauli blocking, their contribution is denoted by "$^0$".

\subsection{Boron}

\begin{table*}[h]
\begin{center}
 \begin{tabular}{|c|c|c|c|c|c|c|c|c|c|c|}
\hline
isotope&feeddown& $A$  &  $Z$ &  $B_{A,Z}/A $&  $ g_{A,Z} $  & $E^{\rm thresh}_{AZ}$ & $E_i\,\, [g_i]$ & $R_{A,Z}^{\rm res.gas}( 1.29)$& $R_{A,Z}^{\rm vir}( 1.29)$ &$R_{A,Z}^{\rm medium}( 1.29)$  \\

\hline
\hline
10B		&-			&10 & 5 & 6.475& 7  & 4.466  & 0.718  [3], 1.740 [1], 2.154 [3]   								&1.363	 	&1.357		&1.086	\\
10B$^0$	&-			&10 & 5 & 6.475&(7) & 4.466  &  3.587 [5]   															&0.0443&0.04117	&0.0324	\\
11B		&-			&11 & 5 & 6.928& 4   &8.674  & 2.124  [2], 4.445 [6], 5.020 [4], 6.741 [8]   				& 1.175	&1.175		&0.9	\\
12B		&-			&12 & 5 & 6.631& 3   &3.369  & 0.953 [5], 1.674 [5]  													&2.251		&2.227		&1.626		\\
12B$^0$	&[$\to ^{11}$B]	&12 & 5 & 6.631& 3   &3.369  &  2.621 [3], 2.723 [1]  							&0.1715	&0.1582	&0.1127	\\
13B		&-			&13 & 5 & 6.496& 4   & 4.879 &3.482 [?], 3.535 [?], 3.681 [?], 3.712 [?], 4.131 [?] 	& 1			&0.9966	&0.6976	\\
14B$^0$	&-			&14 & 5 & 6.102& 5   & 0.970 &- 																				& 1			&0.9341	&0.6065	\\
14B$^0$	&[$\to ^{13}$B]	&14 & 5 & 6.102&(5) & 0.970 &0.740 [3] 													& 0.3381	&0.3001	&0.188	\\
14B*		&$\to ^{13}$B	&14 & 5 & 6.102&(5)  & 0.970 & 1.38 [7]													& 0.4803	&0.387		&0.1801	\\
15B		&-			&15 & 5 & 5.880& 4   & 2.78   & (?)																				& 1			&0.983		&0.6246	\\
16B*		&$\to ^{15}$B	&16 & 5 & 5.507& 1   &-0.082 & (?)															& 1			&0.8574	&0.4311\\
17B		&-			&17 & 5 & 5.270& 4   & 1.39   & (?)																				& 1			&0.9515	&0.5399	\\
18B*		&$\to ^{17}$B	&18 & 5 & 4.977& 5   &-0.005 & (?)															& 1			&0.8659	&0.3832	\\
\hline
 \end{tabular}
\caption{Data of  boron nuclei [units: MeV, fm]. 
Mass number  $A$, charge number $Z$. Ground state binding energy $B_{A,Z}$ and degeneracy $g_{A,Z}$, excitation energy $E_i$ and degeneracy $g_i$ as well as continuum threshold energy $E^{\rm thresh}_{AZ}$ according \cite{Nudat}.   Excited state multiplier $R_{A,Z}^{\rm excited}(\lambda_T)$ for $\lambda_T=1.29$ MeV. }
\label{tab:B}
\end{center}
\end{table*}

Results for B isotopes are shown in Tab. \ref{tab:B}.
As before, we denoted the weakly bound part of the intrinsic partition function with $^{10}$B$^0$, $^{12}$B$^0$, and $^{14}$B$^0$. 
$^{16}$B feeds $^{15}$B, and $^{18}$B feeds $^{17}$B.
It is expected that continuum correlations and in-medium effects may give further corrections to describe the final yields of the isotopes.

\subsection{Carbon}

\begin{table*}[h]
\begin{center}
 \begin{tabular}{|c|c|c|c|c|c|c|c|c|c|c|}
\hline
isotope&feeddown& $A$  &  $Z$ &  $B_{A,Z}/A $&  $ g_{A,Z} $  & $E^{\rm thresh}_{AZ}$ & $E_i\,\, [g_i]$ & $R_{A,Z}^{\rm res.gas}(1.29)$& $R_{A,Z}^{\rm vir}(1.29)$ &$R_{A,Z}^{\rm medium}( 1.29)$  \\
\hline
13C		&-			&13 & 6 & 7.469 & 2 & 4.946 & 3.089 [2], 3.684 [4], 3.853	[6]										& 1.358	 	&1.353		&0.9905	\\
14C		&-			&14 & 6 & 7.520 & 1 & 8.176 & 6.093 [3], 6.589 [7]														& 1.069		&1.069		&0.7491	\\
15C		&-			&15 & 6 & 7.100 & 2 & 1.218 &-																						& 1	 			&0.945		&0.6169	\\
15C$^0$	&[$\to ^{14}$C]&15 & 6 & 7.100  &(2)& 1.218 & 0.740 [6]													&1.69 			&1.532		&0.9775	\\
15C*		&$\to ^{14}$C	&15 & 6 & 7.100 &(2)& 1.218 &3.103 [2], 4.220 [6], 4.657 [4], 4.780 [4]	& 0.3074		&0.004482&0.000206	\\
16C		&-			&16 & 6 & 6.922 & 1 & 4.250 & 1.766 [5]																			&2.272	 		&2.26		&1.445\\
16C$^0$	&[$\to ^{15}$C]	&16 & 6 & 6.922 &(1)& 4.250 & 3.986 [5], 4.089 [7], 4.142 [9]					&0.885		&0.7875	&0.4683	\\
17C$^0$	&[$\to ^{16}$C]	&17 & 6 & 6.558 & 4 & 0.734 & - 																	&1				&0.9218	&0.5379	\\
17C$^0$	&[$\to ^{16}$C]	&17 & 6 & 6.558 &(4)& 0.734 & 0.217 [2], 0.332 [6] 									&1.582			&1.438		&0.8288	\\
17C*		&$\to ^{16}$C	&17 & 6 & 6.558 & (4)& 0.734 & 2.15 [8], 2.71 [2], 3.085 [10] 					& 0.6677 		&0.09067	&0.002612	\\
18C		&-			&18 & 6 & 6.426 & 1 & 4.18   & 1.588 [5], 2.515 [5]														& 3.172		&3.153		&1.843	\\
19C$^0$	&[$\to ^{18}$C]	&19 & 6 & 6.118 & 2 & 0.580 & 0.209 [4], 0.282 [6]									& 5.112	  	&4.665		&2.431	\\
19C*		&$\to ^{18}$C	&19 & 6 & 6.118 &(2)& 0.580 & 0.653 [6], 1.46 [6]										& 2.776	 	&2.383		&0.9985	\\
20C		&-			&20 & 6 & 5.961 & 1 & 2.98   & 1.618 [5]    																		& 2.426	  	&2.391		&1.268		\\
\hline
 \end{tabular}
\caption{Data of  carbon nuclei [units: MeV, fm]. 
Mass number  $A$, charge number $Z$. Ground state binding energy $B_{A,Z}$ and degeneracy $g_{A,Z}$, excitation energy $E_i$ and degeneracy $g_i$ as well as continuum threshold energy $E^{\rm thresh}_{AZ}$ according \cite{Nudat}.  Excited state multiplier $R_{A,Z}^{\rm vir}(\lambda_T)$ for $\lambda_T=1.29$ MeV.}
\label{tab:C}
\end{center}
\end{table*}

For carbon, several states are close to the edge of the continuum, see Tab. \ref{tab:C}. In particular, $^{19}$C is weakly bound, as is $^{17}$C, with the threshold of the continuum below 1 MeV. We expect that these weakly bound nuclei are stronger influenced by in-medium modifications, in particular Pauli blocking.

\section{Estimate for the Pauli blocking shift}
\label{Paulibl}
There are two contributions to the in-medium shift of quasiparticles, the self-energy contribution and the Pauli blocking effects. 
Self-energy shifts are calculated, e.g., by the relativistic mean-field approximation. If the momentum dependence of the shift is neglected,
each nucleon suffers the same shift so that it can be implemented in the chemical potential. The composition is not influenced. For a more detailed discussion see \cite{R09,R11}. 

We give an estimate for the Pauli blocking $\Delta E^{\rm Pauli} _{A,Z}({\bf P};T,n_B,Y_p)$ as function of total momentum $\hbar {\bf P}$ and the
thermodynamic variables $T,\mu_n,\mu_p$ respectively $T,n_B,Y_p$ as
\begin{eqnarray}
\Delta E^{\rm Pauli} _{A,Z}({\bf P};T,n_B,Y_p)&=&e^{-\frac{\hbar^2 P^2}{2 A^2mT}} 2\,(Nn_n+Zn_p)\nonumber\\
&& \times a_{A,Z} e^{-b_{A,Z} T}
\end{eqnarray}
according to Eq. (21) in Ref. \cite{R20}, where also values for the parameters $a_{A,Z}, b_{A,Z}$ are given. 
An average value for $10 \le A \le 16$ is $\bar a = 532.0$ MeV fm$^3$ and $\bar b = 0.05103$ MeV$^{-1}$.
The $\bf P$ dependence is according Eq. (43) of Ref. \cite{R09}.

Within an exploratory calculation, we use the average value to estimate the Pauli blocking shift.
Values for $\Delta E^{\rm Pauli} _{A,Z}({\bf P}=0;T,n_B,Y_p)=\Delta E^{\rm Pauli} _{A,Z}(0)$ are shown in Tabs. \ref{241PuG}, \ref{241PuH}.
The Mott condition where the Pauli blocking shift exceeds the binding energy is fulfilled for $^{11}$Li and $^{19}$C,
but also for excited states of $^{11}$Be, $^{14}$B, $^{15}$C, $^{16}$C, and $^{17}$C. 
Further ground states and excited states, denoted by "$^0$",
are also approaching the continuum edge and partially suppressed.

The Mott condition does not mean that all clusters of the corresponding state are dissolved. The transition to the continuum 
gives also a contribution like a resonance, see the case of $^2$H. An important point is that the Pauli blocking depends on the 
 total momentum $\hbar {\bf P}$, see \cite{R11}.
For $\Delta E^{\rm Pauli} _{A,Z}(0) \ge B_{A,Z}$, the Mott momentum follows as 
\begin{equation}
P^{\rm Mott}= \left\{2 A^2 \frac{T m}{\hbar^2} \ln \left[\frac{\Delta E^{\rm Pauli} _{A,Z}(0)}{B_{A,Z}}\right]\right\}^{1/2}.
\end{equation}

Not only the high-momentum bound states but also part of the continuum states may evolve to the ground state in the final distribution. 
However, this leads to the discussion of branching ratios for the reaction network describing the evolution from freeze-out to the final observed yields,
what is beyond the scope of the present work.

\section{Evaluation of the intrinsic partition function multipliers $R$: Virial approximation}
\label{sec:vir}

In (1), the intrinsic partition function multiplier was introduced as a correction to the simple ideal gas approximation where for each isotope $\{A,Z\}$ only the ground state was considered. 
However, for each channel we have to consider the intrinsic partition function, in particular the account of all excited states. The resonance gas approximation (2) considers all excited, including unstable, states, also resonances in the continuum. 

A systematic description of correlations in the continuum is given by the virial approximation. 
We repeat: we perform a cluster expansion for the interacting system \cite{R20}, and partial densities of different clusters $\{A,Z\}$ are introduced. 
The relevant yields $Y^{\rm rel, vir}_{A,Z}$ in the virial approximation are calculated as 
\begin{eqnarray} 
\label{Y0}
Y^{\rm rel, vir}_{A,Z} &\propto & g_{A,Z} \left(\frac{2 \pi \hbar^2}{A m \lambda_T}\right)^{-3/2} \times  \nonumber \\&&
e^{(B_{A,Z}+(A-Z) \lambda_n+Z  \lambda_p)/ \lambda_T}\, R^{\rm vir}_{A,Z}(\lambda_T)
\end{eqnarray}  
(nondegenerate limit), where $B_{A,Z}$ denotes the (ground state) binding energy and $g_{A,Z}$ the degeneracy \cite{Nudat}.
The factor 
\begin{eqnarray}
\label{Rfactor}
R^{\rm vir}_{A,Z}(\lambda_T)=1+\sum^{\rm exc}_i [g_{AZ,i}/g_{A,Z}] e^{-E_{AZ,i}/\lambda_T}
\end{eqnarray}
is related to the intrinsic partition function of the cluster $\{A,Z\}$. The index $i$ characterizes further quantum numbers of the excited state such as angular momentum.
The summation is performed over all excited states of excitation energy $E_{AZ,i}$ and degeneracy $g_{AZ,i}$  \cite{Nudat}.

As already discussed for H, He, we have also to consider the scattering states. This leads to a change in the contributions obtained for bound states, 
and, in particular, for the unbound states. We use the  Beth-Uhlenbeck relation, see \cite{rnp20} to perform the sum over the continuum states,
\begin{eqnarray}
\label{Rviri}
&&\!\!\!\!\!\!C^{\rm vir}_{AZ,i}(\lambda_T)=[1-e^{-(E^{\rm thresh}_{A,Z}-E_{AZ,i} )/\lambda_T}]\Theta(E^{\rm thresh}_{A,Z}-E_{AZ,i} ) \nonumber \\
&&\!\!\!\!\!\!+\frac{1}{\pi \lambda_T}
e^{-(E^{\rm thresh}_{A,Z}-E_{AZ,i} )/\lambda_T} \int_0^\infty dE e^{-E/\lambda_T}\delta_{AZ,i}(E)) ,
\end{eqnarray}
as prefactor for the contribution of the different channels $\{A,Z,i\}$, degeneracy $g_{AZ,i}$. 
$E^{\rm thresh}_{A,Z}$ is the difference of the binding energy of the ground state and the binding energy of the components of the lowest continuum, 
the $\Theta$ function is 1 if a bound state occurs. 
 If a channel has no common bound state, we have $E_{AZ,i}=0$,
 all excited states are in the continuum. 
  The intrinsic partition function multiplier $R^{\rm vir}_{A,Z}(\lambda_T)$ (\ref{Rfactor}) can be rewritten as 
\begin{eqnarray}
\label{Rfactor1}
\!\!\!R^{\rm vir}_{A,Z}(\lambda_T)=\sum^{\rm exc}_i \frac{g_{AZ,i}}{g_{A,Z}} e^{-E_{AZ,i}/\lambda_T}C_{A,Z}^{\rm vir}(\lambda_T,E_{AZ,i}).
\end{eqnarray}

As in the case of the resonance gas approximation, the sum over all states $i$ can be subdivided into different contributions $\alpha$, 
\begin{eqnarray}
\label{Rfactor2}
R^{\rm vir}_{A,Z}(\lambda_T)=\sum_\alpha R^{{\rm vir},\alpha}_{A,Z}(\lambda_T).
\end{eqnarray}
We divided it into the contribution of well-bound nuclear states with binding energy larger than 1 MeV ($R$) and weakly bound states, ($R^0$), both decaying to the ground state, and in unbound states ($R^*$) which feed down to other channels.
The same subdivision has been performed also for the intrinsic partition function multiplier $R^{\rm res.gas}_{A,Z}(\lambda_T)$. We perform this subdivision to calculate the feed-down process from freeze-out to the final yields.
 
 We are not able here to evaluate the integral over the scattering phase shifts for all channels of interest. We give only an estimate
 in analogy of the cases $^2$H, $^4$H, $^5$He which have been discussed in Ref. \cite{R20}. 
The relation  $R^{\rm vir}_{AZ}(T)=e^{(E^{\rm thresh}_{A,Z}-E^{\rm eff}(T))/T}$ was given there.
 We use the effective energy $E^{\rm eff}(T)$ given there and have for $^4$H ($S_n=-1.6$ MeV) the value 
 $R^{\rm vir}_{^4{\rm H}}(1.29)=0.057$, and  for $^5$He for $S_n=-0.735$ MeV the value 
 $R^{\rm vir}_{^5{\rm He}}(1.29)=0.731$. Together with the known virial coefficient in the deuteron channel, 
in \cite{rnp20} the interpolation formula (in units of MeV)
\begin{eqnarray}
\label{fitRvir}
C_{A,Z}^{\rm vir}(\lambda_T,E_{AZ,i})&=&\frac{1}{e^{-(E^{\rm thresh}_{A,Z}-E_{AZ,i}+1.129)/0.204}+1} \times \nonumber \\
&&\frac{1}{e^{-(E^{\rm thresh}_{A,Z}-E_{AZ,i}+2.45)/\lambda_T}+1}
\end{eqnarray}
has been introduced which reproduces the values for $^2$H, $^4$H, and $^5$He at $\lambda_T=1.29$ MeV. 
If we have multiple contributions within a group $\alpha$ we replace $C_{AZ,i}^{\rm vir}(\lambda_T,E_{AZ,i})$ by its value for the lowest excitation energy $E_{AZ,\alpha}$ of the group,
and can extract this factor from the summation so that 
\begin{eqnarray}
\label{Rfactor3}
R^{{\rm vir}, \alpha}_{A,Z}(\lambda_T)=C_{A,Z}^{\rm vir}(\lambda_T,E_{AZ,\alpha})R^{{\rm res.gas},\alpha}_{A,Z}(\lambda_T).
\end{eqnarray}
We used this expression to calculate the multipliers $R_{A,Z}^{{\rm vir},\alpha}(\lambda_T)$ shown in the Tables.

\section{Evaluation of the intrinsic partition function multipliers $R$: Medium modifications}

The virial expression uses the bound state energies and scattering phase shifts of the free clusters, neglecting interaction effects. 
The most important in-medium effects are self-energy and Pauli blocking.
Self-energy effects are well known and are parametrized, for instance, as Skyrme forces or relativistic mean-field approximation. 
If we neglect the momentum dependence of the single-particle mean-field shift, this self-energy shift $\Delta^{\rm SE}_\tau$ for neutron and protons, 
which occurs also in the bound states may be included in the effective chemical potentials $\tilde \lambda_\tau=\lambda_\tau -\Delta^{\rm SE}_\tau$. 
Values for $\Delta^{\rm SE}_\tau$ have been given in Sec. \ref{inmedium}.
Thus, it has no influence on the composition at given temperature and densities.

In contrast, the Pauli blocking shift occurs only for interacting nucleons in a cluster and depends strongly on the quantum state, as discussed above. 
Therefore we take this shift of energy levels explicitly into account. In principle, also the scattering phase shifts are modified. See \cite{SRS} for the two-nucleon case, where a generalized Beth-Uhlenbeck approach is given which accounts for in-medium effects.

The Pauli blocking has different consequences. (i) In (\ref{Y0}), (\ref{Rfactor}), 
the energy shift of the bound states $\Delta E^{\rm Pauli}_{A,Z,i}$ 
has to be added to the binding energies $B^{\rm medium}_{A,Z}=B_{A,Z}-\Delta E^{\rm Pauli}_{A,Z}$ for the ground state, 
but also for all  excitation energies $E^{\rm medium}_{AZ,i}=E_{AZ,i}+\Delta E^{\rm Pauli}_{AZ,i}-\Delta E^{\rm Pauli}_{A,Z}$ which are taken relatively to the ground state. 
In principle, the Pauli blocking depends on the wave function of the bound state as well as the total momentum $\bf P$ of the cluster, 
but for simplicity we assume that we can neglect this and approximate these different Pauli blocking shifts by $\Delta E^{\rm Pauli}_{A,Z}$. 
Then, a common factor can be extracted so that $e^{-\Delta E^{\rm Pauli}_{A,Z}/\lambda_T}R_{A,Z}^{\rm vir}(\lambda_T)$ appears. 

(ii): The medium modifications will also result in a density dependence of the intrinsic partition function or the corresponding factors  $R^{{\rm medium}, \alpha}_{A,Z}(T,n_B,Y_p)$ in the decomposition  corresponding to Eq. (\ref{Rfactor3}).
We use the virial expression (\ref{Rfactor3}) but replace in $C_{A,Z}^{\rm vir}(\lambda_T,E_{AZ,\alpha})$ the excitation energy $E_{AZ,\alpha}$ by $E_{AZ,\alpha}+\Delta E^{\rm Pauli}_{A,Z}$,
\begin{eqnarray}
\label{Rfactor4}
R^{{\rm medium}, \alpha}_{A,Z}(\lambda_T,\lambda_n,\lambda_p)&&=C_{AZ}^{\rm vir}(\lambda_T,E_{AZ,\alpha}+\Delta E^{\rm Pauli}_{A,Z}) \times \nonumber \\
&&R^{{\rm res.gas},\alpha}_{A,Z}(\lambda_T)e^{-\Delta E^{\rm Pauli}_{A,Z}/\lambda_T}.
\end{eqnarray}

The shift of the bound states gives a decrease of the binding energy which may disappear (Mott effect). The dissolution of bound states opens additional channels for feed-down processes. 

We discuss three different cases: Well bound states with binding energy larger than $T\approx 1$ MeV are shifted but not dissolved. 
Because the temperature is low, the effect of continuum correlations and the threshold energy is small for the strongly bound isotopes so that, in addition to the self-energy shift $\Delta^{\rm SE}_{A,Z}$, 
the factor $ \exp[-\Delta E_{A,Z}^{\rm Pauli}/\lambda_T]$ determines the influence of the medium.
The contribution of scattering states has been approximated by Eq. (\ref{fitRvir}), replacing $E_{AZ,i}$ by $E_{AZ,i}+\Delta E^{\rm Pauli}_{AZ,i}$.

Of special interest are the weakly bound states where the shift brings it above the continuum edge, $E_{AZ,i}+\Delta E^{\rm Pauli}_{A,Z}-E^{\rm thresh}_{AZ} >0$ 
so that the bound state will be dissolved and feed down other final states.  
We estimate these contributions also according to Eq. (\ref{fitRvir}) as done above for the scattering states. 
Instead to feed down the ground state after de-excitation, we assume that they feed down other isotopes as indicated in Tabs. \ref{241PuG}, \ref{241PuH}.

This simplified picture to describe the reaction processes after freeze-out has to be improved in future treatments within a systematic nonequilibrium approach.
As known from reaction networks describing expanding matter in astrophysics, branching ratios may be introduced to describe the evolution of the primary distribution to the final distribution,
which allow for different final states of an excited state. 
For our considerations this is not of relevance as long as the contribution of excited states (including continuum states) is small,
and the main contribution to the yield of isotopes is determined by the strongly bound states. 
Another situation appears for the weakly bound final states like $^{11}$Li and $^{19}$C in our calculation, see Tabs. \ref{241PuG}, \ref{241PuH}.
Because they are shifted to the continuum, in our simple treatment of the feed down process they will feed other isotopes, and the final yield for both isotopes will be zero.
A more detailed description would allow that some of the primary states may remain in the bound state. 
For instance, the Pauli blocking is momentum dependent, and clusters  with large momentum remain bound because Pauli blocking is effective only within the Fermi sphere, as discussed in Sec. \ref{Paulibl} and Sec. \ref{inmedium}.
There is also a finite probability that continuum contributions according to the Beth-Uhlenbeck formula will de-excite to the ground state according to a coalescence model.
For this one has to introduce branching ratios which allow feed-down to different final states. 
Even being small, special feed-down processes become visible if they are the only process to form a final bound state.

We are not able in this work to derive a microscopic expression for the branching ratios during expansion after freeze-out. 
They are of relevance when they are the only process to populate the final yield, for instance, of the dissolved isotopes $^{11}$Li and $^{19}$C in Tabs.  \ref{241PuG}, \ref{241PuH}.
To give an example, we assumed in both cases a branching ratio 0.1 that the primary yield remains at the same isotope to form the final yield. 
As shown in Fig. \ref{fig:medium}, a branching ratio would nicely fit the general behavior of the calculated yields.
This example is only for illustration, the derivation of the branching ratio  goes beyond the frame of the present work.


Another issue which needs further considerations is the threshold energy $E^{\rm thresh}_{A,Z}$ which is taken in this work from the data tables.
It is possible that this quantity is also changed in a dense medium. 
For this, a nonequilibrium approach to the reaction processes is necessary  which finally will provide us with in-medium branching ratios discussed above.
Our description of the reactions during the expansion process after freeze-out remains a semi-empirical one with a simplified reaction network as shown in Tabs. \ref{241PuG}, \ref{241PuH}.

%
%

 \begin{table*}
\begin{center}
\hspace{0.5cm}
 \begin{tabular}{|c|c|c|c|c|c|c|c|c|c|c|c|}
\hline
isotope& $A$  &  $Z$ &  $\frac{B_{A,Z}}{A} $&  $ g_{A,Z} $  & $E^{\rm thresh}_{AZ}$ & $R_{A,Z}^{\rm vir}(1.29)$  & $\Delta E^{\rm Pauli} _{A,Z}$ [MeV]
& $R_{A,Z}^{\rm medium}(1.29)$&$Y_{A,Z}^{\rm rel,medium}$ &$Y_{A,Z}^{\rm final,medium}$ &$Y^{\rm obs}_{A,Z}$/$Y^{\rm final,medium}_{A,Z}$\\
\hline
1n		&1&0	&0		& 2 		&-			&1 			&-      		&1      		&0.11203   	&0.11203 			& 0.9551\\
1H		&1& 1 	&0		& 2		&-              		&1              	&-   		 	&1      		&4.563E-6    	 &4.563E-6		&-	\\
2H 		&2& 1 	& 1.112 	& 3 		&2.224		& 0.9739		&0.057    		&0.9306   		&9.226E-6 	&9.226E-6  		&0.9173\\
3H		&3& 1 	& 2.827 	& 2 		&6.257		& 0.9988		&0.114    		&0.9138    	&1.289E-4   	&1.342E-4 		&1.223 \\
4H$^*$	&4& 1 	& 1.720	& 5 		&-1.6		&0.09404		&0.172     		& 0.03564 	&5.146E-6  	 & [$\to ^3$H]   		 &-  \\
\hline
3He		&3& 2 	& 2.573	& 2 		& 5.494		& 0.9979		&0.057     		&0.9545      	& 2.948E-9	& 2.948E-9   		&- \\
4He		&4& 2 	& 7.073 	& 1 		& 20.577		& 1 			&0.114    		&0.915   		&1.639E-3  	&1.935E-3		& 1.042\\
5He$^*$	&5& 2 	& 5.512	& 4 		&-0.735		& 0.6906 		&0.172    		& 0.5029 		&2.622E-4   	&[$\to ^4$He] 		&-\\
6He$^0$	&6& 2 	& 4.878 	& 1 		& 0.975		& 0.9343 		& 0.2291   	&0.7723    	&4.575E-5   	&5.884E-5		&0.8903\\
6He$^*$	&6& 2 	& 4.878 	& 1 		& 0.975		& 0.7919 		& 0.2291   	& 0.4617   	&2.735E-5   	&[$\to ^4$He] 		&-\\
7He$^*$	&7& 2 	& 4.123	& 4		& -0.410		&0.9313		& 0.2864    	&0.6577 		&1.309E-5  	&[$\to ^6$He]		& -\\
8He		&8& 2 	& 3.925 	& 1 		& 2.125		& 0.972 		& 0.3437  		&0.7383 		&2.919E-6  	&2.935E-6		&1.029\\
8He$^*$	&8& 2 	& 3.925 	& 1 		& 2.125		& 0.2332		& 0.3437  		&0.06923   	&2.737E-7  	&[$\to ^4$He]		& -\\
9He$^*$	&9& 2 	& 3.349	& 2 		& -1.25		& 0.364 		& 0.4009  		& 0.04882  	&1.605E-8   	 &[$\to ^8$He]		&-\\
\hline
6Li		&6 & 3 	& 5.332	& 3 		& 1.475     	& 0.9545    	& 0.172    		&0.83    		& 4.78E-8    	& 4.78E-8			&-\\
6Li$^*$	&6 & 3 	& 5.332	& 3 		& 1.475     	& 0.3847    	& 0.172    		& 0.2842     	& 1.417E-8    	&[$\to ^3$H]		&-\\
7Li		&7 & 3 	& 5.606	& 4 		& 2.461		& 1.3159		& 0.2291    	&1.0972  		&2.66E-6 		&2.985E-6		&0.4523  \\
8Li		&8 & 3 	& 5.160 	& 5 		& 2.038     	& 1.2424		& 0.2864    	&0.9877     	&1.624E-6   	&1.624E-6  		&0.5211\\
8Li$^*$	&8 & 3 	& 5.160 	&(5) 		& 2.038     	& 0.2647		&  0.2864   	&0.1975		& 3.247E-7  	&[$\to ^7$Li]		& -\\
9Li		&9 & 3 	& 5.038 	& 4 		& 4.062     	& 1.0553 		& 0.3437  		& 0.8069   	&2.694E-6 	&2.876E-6 		&0.5814\\
10Li$^*$	&10& 3 	& 4.531	&{\it 3}	& -0.032 		& 0.863    		& 0.4009  		&0.5866		&1.539E-7   	&[$\to ^9$Li]		&-\\
11Li$^+$	&11& 3	& 4.155 	& 4 		&{\bf 0.396}	& 0.9003		&{\bf 0.4582}	& 0.6026		&3.03E-8$\times 0.1$ 	&3.03E-9 		&0.2992\\
11Li$^0$	&11& 3	& 4.155 	& 4 		&{\bf 0.396}	& 0.9003		&{\bf 0.4582}	& 0.6026		&3.03E-8$\times 0.9$	&[$\to ^9$Li]	 &-\\
12Li$^*$	&12& 3	& 3.792 	&{\it 3}	& -0.201 		& 0.8422 		& 0.5155   	&0.4697		&1.585E-9	&[$\to ^{9}$Li] 		&-\\
\hline
7Be		&7  & 4 	& 5.372	& 4 		& 1.585 		& 1.3015		& 0.172    		&1.1324   		&3.058E-11 	&3.058E-11     		&-\\
8Be$^*$	&8  & 4 	& 7.062	& 1 		& -0.088 		& 1.2657		&0.2291    	&1.0192		& 1.722E-6      	& [$\to ^4$He] 		&-\\
9Be		&9  & 4 	& 6.462	& 4 		& 1.558  		& 0.9572		& 0.2864    	&0.7585  		&2.03E-6 		&2.778E-6 		&0.3191 \\
9Be$^*$	&9  & 4 	& 6.462	& 4 		& 1.558  		& 0.4796		& 0.2864    	&0.3629   		&9.712E-7 	& [$\to ^4$He] 	 	&- \\
10Be	&10 & 4 	& 6.497 	& 1 		& 6.497  		& 1.366		& 0.3437   	&1.0462    	&1.464E-5 	&1.752E-5 		&0.5285 \\
10Be$^0$	&10 & 4 	& 6.497 	& (1) 	& 6.497  		& 0.07181		& 0.3437   	& 0.05348		& 7.482E-7    	&[$\to ^9$Be]  		&-\\
11Be	&11 & 4	& 5.953 	& 2 		& 0.502  		& 0.9076		& 0.4009  		&0.6422  		&2.829E-6	&2.829E-6 		&0.4202 \\
11Be$^0$	&11 & 4	& 5.953 	& 2 		& 0.502  		& 0.6894		& 0.4009  		&0.48015 		&1.141E-6	&[$\to ^{10}$Be]	&- \\
11Be$^*$	&11 & 4	& 5.953 	& (2) 	& 0.502  		& 0.308		& 0.4009   	&0.03957		& 3.341E-9    	& [$\to ^{10}$Be]	& -\\
12Be	&12 & 4	& 5.721 	& 1 		& 3.170  		& 2.1223		& 0.4582   	&1.4798   		&3.957E-6   	&4.074E-6  		&0.1385 \\
12Be$^0$	&12 & 4	& 5.721 	& (1) 	&3.170  		& 0.3307		& 0.4582   	&0.222		&5.937E-7 	& [$\to ^{10}$Be]	& -\\
13Be$^*$	&13 & 4	& 5.241 	& 2 		& -0.510 		& 0.7806		& 0.5155   	&0.3144		& 1.164E-7	& [$\to ^{12}$Be] 	& -\\
14Be	&14 & 4	& 4.994 	& 1 		& 1.264  		& 0.9468		& 0.5728   	&0.5897  		&4.497E-8  	&4.498E-8 		&0.01209 \\
15Be$^*$	&15 & 4	& 4.541 	& 6 		& -1.800 		& 0.02241		&0.6301   		&0.0005243	&5.994E-12	& [$\to ^{14}$Be]	& -\\
\hline
 \end{tabular}
\caption{Calculated yields per fission of ternary fission of $^{241}$Pu($n_{\rm th}$,f), in-medium effects included. 
For the fit of the $^{241}$Pu($n_{\rm th}$,f) data, the  Lagrange parameter values are
$\lambda_T=1.29,\,\,\tilde \lambda_n=-3.09 ,\,\tilde \lambda_p=-16.19$ MeV,  volume 1796 fm$^3$, fit metric 0.0095 for all nuclei $Z \le 2$. 
Baryon density $n_B=6.7 \times 10^{-5}$, $Y_p=0.035$.
Bold figures: Mott effect, the Pauli blocking shift exceeds the ground state binding energy so that the bound state is dissolved. 
For illstration of this effect, a branching ratio 0.1 is assumed to remain as final yield (denoted by superscript "+") whereas 0.9 is assumed to feed down the yields of other isotopes.}
\label{241PuG}
\end{center}
\end{table*}

 \begin{table*}
\begin{center}
\hspace{0.5cm}
 \begin{tabular}{|c|c|c|c|c|c|c|c|c|c|c|c|}
\hline
isotope& $A$  &  $Z$ &  $\frac{B_{A,Z}}{A} $&  $ g_{A,Z} $  & $E^{\rm thresh}_{AZ}$ & $R_{A,Z}^{\rm vir}(1.29)$  & $\Delta E^{\rm Pauli} _{A,Z}$ [MeV]
& $R_{A,Z}^{\rm medium}(1.29)$&$Y_{A,Z}^{\rm rel,medium}$ &$Y_{A,Z}^{\rm final,medium}$ &$Y^{\rm obs}_{A,Z}$/$Y^{\rm final,medium}_{A,Z}$\\
\hline
10B	&10 & 5 & 6.475 & 7   		& 4.466	& 1.357		& 0.2864    	& 1.086		&3.547E-9 	& 3.547E-9   	&- \\
11B	&11 & 5 & 6.928 & 4  		& 8.674	& 1.175		& 0.3437   	& 0.9   		& 1.261E-6   	&1.428E-6	 & 0.2258\\
12B	&12 & 5 & 6.631 & 3  		& 3.369  	& 2.227 		& 0.4009   	& 1.626 		&2.411E-6   	&2.411E-6 	 &0.08358\\
12B$^0$	&12 & 5 & 6.631&(3)  	& 3.369  	& 0.1582		& 0.4009   	& 0.1127 		& 1.671E-7         &  [$\to ^{11}$B] & -\\
13B	&13 & 5 & 6.496	& 4   		& 4.879 	& 0.9966		& 0.4582    	& 0.6976    	& 6.209E-6         &7.096E-6 	& -\\
14B	&14 & 5 & 6.102 & 5   		& 0.970 	& 0.9341		& 0.5155    	&0.6065     	&1.461E-6  	&1.461E-6  	&0.01793\\
14B$^0$	&14 & 5 & 6.102 & (5)   	& 0.970 	& 0.3001		& 0.5155    	&0.188    		&4.53E-7  	&[$\to ^{13}$B]  & -\\
14B$^*$	&14 & 5 & 6.102 & (5)   	& 0.970 	& 0.387		& 0.5155    	&0.1801		&4.338E-7   	&  [$\to ^{13}$B]  & -\\
15B	&15 & 5 & 5.880 & 4   		& 2.78 	& 0.9829		& 0.5728   	&0.6246 		&1.058E-6   	&1.075E-6 	& 0.008624\\
16B$^*$	&16 & 5 & 5.507	& 1   & -0.082 	& 0.8574		& 0.6301  		&0.4311 		& 1.723E-8  	 & [$\to ^{15}$B]   & -\\
17B	&17 & 5 & 5.270 & 4   		& 1.39 	& 0.9515		&0.6873   		&0.5399		&2.723E-8	& 2.964E-8 	& -\\
18B$^*$	&18 & 5 & 4.977	& 5   & -0.005	& 0.8659	 	& 0.7446  		&0.3832		& 2.407E-9        &  [$\to ^{17}$B]& -\\
\hline
13C	&13 & 6 & 7.470 & 2 			& 4.946	& 1.353 		& 0.4009   	& 0.9905 		&3.134E-6        &3.137E-6  	& -\\
14C	&14 & 6 & 7.520 & 1 			& 8.176	& 1.069 		& 0.4582  		& 0.7491		&6.769E-5	 &1.139E-4 	&0.02229 \\
15C	&15 & 6 & 7.100 & 2 			& 1.218	& 0.9449		&0.5155    	&0.6169  		&2.916E-5  	&5.92E-5		&0.01464 \\
15C$^0$	&15 & 6 & 7.100 & (2) 	& 1.218	& 1.5315		&0.5155    	&0.9775    	&4.621E-5  	& [$\to ^{14}$C] &- \\
15C$^*$	&15 & 6 & 7.100 & (2) 	& 1.218	& 0.004482	& 0.5155   	& 0.000206	&9.728E-9         & [$\to ^{14}$C] & -\\
16C	&16 & 6 & 6.922 & 1 			& 4.250	& 2.259		& 0.5728   	& 1.445 		&9.266E-5 	&1.305E-4 	&0.007721\\
16C$^0$	&16 & 6 & 6.922 & (1) 	& 4.250	& 0.7875		& 0.5728   	& 0.4683		& 3.004E-5	   & [$\to ^{15}$C]    & -   \\
17C	&17 & 6 & 6.558 & 4 			& 0.734 	& 0.9218 		& 0.6301    	&0.5379    	&2.447E-5  	&2.447E-5 	 & 0.005271\\
17C$^0$	&17 & 6 & 6.558 & (4) 	& 0.734 	& 1.438 		& 0.6301   	& 0.8288 		&3.77E-5	  	 &[$\to ^{16}$C]   & -  \\
17C$^*$	&17 & 6 & 6.558 & (4) 	& 0.734 	& 0.09067 	& 0.6301   	& 0.002612 	&1.188E-7	   &[$\to ^{16}$C]   & -  \\
18C	&18 & 6 & 6.426 & 1	 		& 4.18  	& 3.153		& 0.6873   	&1.843		&5.329E-5   	&8.188E-5 	&0.0006891\\
19C$^+$	&19 & 6 & 6.118 & 2 	& {\bf 0.580} 	& 4.665		& {\bf 0.7446}    &  2.431    	&  2.181E-5$\times 0.1$ & 2.181E-6 &0.0002309 \\
19C$^0$	&19 & 6 & 6.118 & 2 	& {\bf 0.580} 	& 4.665		& {\bf 0.7446}    &  2.431    	& 2.181E-5$\times 0.9$ &  [$\to ^{18}$C]   & -    \\
19C$^*$	&19 & 6 & 6.118 & (2) 	& 0.580 	& 2.383		& 0.7446   	&0.9985 		&6.858E-6	   & [$\to ^{18}$C]   & -    \\
20C	&20 & 6 & 5.961 & 1 			& 2.98	& 2.391		& 0.8019   	&1.268 		& 5.646E-6  	&5.646E-6 	&0.0001285\\
\hline
 \end{tabular}
\caption{Continuation of Tab. \ref{241PuG}.  }
\label{241PuH}
\end{center}
\end{table*}

\end{document}